\documentclass[a4,
superscriptaddress,
preprintnumbers,
nofootinbib,
nobibnotes,
amsmath,amssymb,
aps,
pra,
]{revtex4}

\usepackage{amsmath}
\usepackage{fdsymbol}

\usepackage{graphicx}
\usepackage{bm}
\usepackage{latexsym}
\usepackage{array}
\usepackage[dvipsnames]{xcolor}
\usepackage[normalem]{ulem}
\usepackage{hyperref}
\usepackage[dvipsnames]{xcolor}

\usepackage{epstopdf, epsfig}

\usepackage{float}

\usepackage{color}
\newcommand{\jon}{\textcolor{black}}

\begin{document}

\preprint{APS/123-QED}

\title{Reconfiguration and oscillations of a vertical, cantilevered-sheet subject to vortex-shedding behind a cylinder}% in a quasi-2D channel flow}

\author{J. John Soundar Jerome}
\email{john-soundar@univ-lyon1.fr}
\homepage{https://lmfa.ec-lyon.fr/spip.php?article378}
\affiliation{Univ Lyon, Universit\'{e} Claude Bernard Lyon $1$, Laboratoire de M\'{e}canique des Fluides et d'Acoustique, CNRS UMR--$5509$, Boulevard $11$ novembre $1918$, F--$69622$ Villeurbanne cedex, Lyon, France.}%

\author{Yohann Bachelier}
\affiliation{Univ Lyon, Universit\'{e} Claude Bernard Lyon $1$, Laboratoire de M\'{e}canique des Fluides et d'Acoustique, CNRS UMR--$5509$, Boulevard $11$ novembre $1918$, F--$69622$ Villeurbanne cedex, Lyon, France.}

\author{Delphine Doppler}
%\email{delphine.doppler@univ-lyon1.fr}
%\homepage{https://lmfa.ec-lyon.fr/spip.php?article347}
\affiliation{Univ Lyon, Universit\'{e} Claude Bernard Lyon $1$, Laboratoire de M\'{e}canique des Fluides et d'Acoustique, CNRS UMR--$5509$, Boulevard $11$ novembre $1918$, F--$69622$ Villeurbanne cedex, Lyon, France.}

\author{Christophe Lehmann}
%\altaffiliation[Also at ]{Institute of Fluid Mechanics, Technische Universit\"{a}t Dresden, George-B\"{a}hr Str. 3c,01062 Dresden, Germany.}
\affiliation{Univ Lyon, Universit\'{e} Claude Bernard Lyon $1$, Laboratoire de M\'{e}canique des Fluides et d'Acoustique, CNRS UMR--$5509$, Boulevard $11$ novembre $1918$, F--$69622$ Villeurbanne cedex, Lyon, France.}
%\affiliation{Institute of Fluid Mechanics, Technische Universit\"{a}t Dresden, George-B\"{a}hr Str.3c, 01062 Dresden, Germany.}%


\author{Nicolas Rivi\`ere}
\affiliation{INSA de Lyon, Laboratoire de M\'{e}canique des Fluides et d'Acoustique, CNRS UMR--$5509$, Boulevard $11$ novembre $1918$, F--$69622$ Villeurbanne cedex, Lyon, France}%

%\collaboration{Institute of Fluid Mechanics, Technische Universit\"{a}t Dresden, George-B\"{a}hr Str. 3c,01062 Dresden, Germany.}%\noaffiliation

\date{\today}% It is always \today, today,

\begin{abstract}
The dynamics of a thin low-density polyethylene sheet subject to periodic forcing due to B\'enard-K\`arm\`an vortices in a long narrow water channel is investigated here. \jon{In particular, the time-averaged sheet deflection and its oscillation amplitude are considered. The former is first illustrated to be well-approximated by the static equilibrium between the buoyancy force, the elastic restoring force, and the profile drag based on the depth-averaged water speed. Our observations also indicate that the presence of upstream vortices hinder the overall reconfiguration effect, well-known in an otherwise steady flow. For the sheet tip oscillations, a simple model based on torsional-spring-mounted flat plate correctly captures the measured tip amplitude $\delta_b$ over a wide range of sheet physical properties and flow conditions.} Furthermore, a rich phenomenology of structural dynamics including vortex-forced-vibration, lock-in with the sheet natural frequency, flow-induced vibration due to the sheet wake, multiple-frequency and modal response is reported.
\end{abstract}

\maketitle

\section{Introduction}
\label{sec:intro}
A rigid circular cylinder, when exposed to a fluid flow, expresses a wide variety of dynamics depending on its mechanical properties, the flow characteristics, and its boundary conditions. Many investigations on the cylinder wake characteristics \cite{Roshko1954_modRe, Roshko1961_highRe, Berger_AnnRevFlu1972, Williamson_AnnRevFlu1996}, and the related \textit{Vortex-Induced-Vibration} (VIV) \cite{Sarpkaya_1979, Bearman_AnnRevFlu1984, Sarpkaya_1995, Sarpkaya_JFS2004, Williamson_2004, Bearman_JFS2011} continue to contribute to a multitude of industrial applications, involving, structural design in both marine and civil engineering, energy harvesting, locomotion, mixing and transfer \cite[among others]{Blevins, Paidoussis_book1998, SummerFredsoe_2006, NaudascherRockwell_book2017}. Often in applications and in nature, mechanical structures that are exposed to a fluid flow could be very flexible in order to alleviate flow-generated forces \cite{Koehl_AmZoo1984, Vogel_1984, Vogel_1989, Koehl_AnnRevEcoSys1996, Vogel_1994, DeLangre2008, Gosselin_JExpBot2019}. The resulting internal bending stresses and the modified flow angle of attack might alter both the structural dynamics and the wake characteristics. Such effects are only recently considered for flow-bent cylinders to explore associated \textit{Fluid-Structure Interactions} (FSI), namely, reduction in oscillation amplitude, multi-frequency response and \textit{lock-in} mode, to name a few \cite[and references therein]{Leclerc_deLangre_2018}. While, a related topic, namely, the flutter of a flat-plate, be it rigid or flexible, had received a large number of attention during the past two decades \cite[for more]{Shelley_AnnRevFlu2011}, rare are the studies on VIV of a single cantilevered slender blades whose one-end is anchored to the flow bed. 
%\textit{Ranunculus penicillatus}, \textit{Fontinalis antipyretica}, \textit{Myriophyllum alterniflorum}, and \textit{Callitriche stagnalis} 

In this context, using single specimens of four different freshwater plants in laboratory flumes whose floor is covered with uniform density of shorter artificial grass, \citet{Siniscalchi_Jhyd2013} correlated individual plant movement and its drag force fluctuations with respect to upstream turbulence. Also, they remarked spatial flapping-like movement in all plant species, with the propagation velocity of perturbations being comparable to the approach flow velocity. More recently, a series of studies by \citet{Jin_PRF2018, Jin_PoF2018, Jin_TandemBlades_JFM2018, Jin_JFM2019} put light upon a rich phenomenology of \textit{Fluid-Structure Interactions} for dynamically \textit{reconfigured} flexible blades. Firstly, \citet{Jin_PRF2018} point out that moderately-long flexible structures (aspect ratio $L_b/w_b \in (5, 25)$) vibrate in the stream-wise direction at their natural frequency while the wake fluctuations beat at frequency of vortex shedding. \citet{Jin_PoF2018} also illustrated that the structural dynamics of moderately-short flexible plates (aspect ratio $L_b/w_b \in (2, 3)$) are governed by both wake-fluctuations and non-linear modulations of structural bending. These cantilevered-structures presented a maximum tip oscillation intensity at some critical Cauchy number which compares the profile drag force experienced by the structure if it were rigid and the characteristic internal restoring force generated due to an external force. More recently, \citet{Jin_JFM2019} used a flexible blade of aspect ratio $4$ at different inclination angles to the incoming flow and highlighted the presence of three modes of tip-oscillations, namely, fluttering, tip-twisting and orbital modes which occur respectively at increasing Cauchy number $C_y$. Orbital modes are characterized by large-amplitude coupled twisting and bending deformations, and they occur for sufficiently large inclination angles. Much is to follow in the perspective of these studies, for instance, the influence of mass ratio, Reynolds and Cauchy numbers on the sheet reconfiguration and sheet tip dynamics such as oscillation amplitudes and frequency content.

The effect of wind on trees and terrestrial plant fields, and of water currents on aquatic vegetation \cite{DeLangre2008, Nepf_AnnRevFlu2012, Gosselin_JExpBot2019} are crucial for sediment transport, water quality and biodiversity of aquatic species. Among the well-known examples, \textit{honami} (Japanese: \textit{ho} $=$ crops and \textit{nami} $=$ wave) \cite{Finnigan_1979honami, Py2006, Gosselin2009} and \textit{monami} (Japanese: \textit{mo} $=$ aquatic plant) \cite{AckermanOkubo_1993monami, Ghisalberti2002, Singh2016, Tschisgale_JFM2021}, respectively, represent coherent motion of crops and aquatic canopies when the flow resistance is sufficiently high. The proposed mechanistic views of such wavy motion generally involve the two-way coupling between flow vortices and the flexible canopy of plants \cite{DeLangre2008, Nepf_AnnRevFlu2012, Nikora_2012} so that velocity spectrum and eddies in the incoming flow modulate the motion of flexible structures \cite{Jin_PRE2016}. Also, there is now some evidence \cite[Chap. $5$]{Barsu_these2016} that the mechanical response of flexible blades in a channel flow might depend strongly on the Cauchy number $C_y$. It seems, therefore, important to know how vortices of different sizes interact with cantilevered flexible blades at various flow velocities and blade physical characteristics.

In the wake of these previous investigations, we propose to study the motion of a thin flexible sheet anchored to the bottom of a water-channel when it encounters a regular array of eddies generated from vortex shedding behind a cylinder (Figure \ref{fig:SchemaManip}). Thereby, we seek to provide a few insights into the  \textit{dynamic} reconfiguration (\ref{sec:SheetReconfig}), tip amplitude (\ref{sec:TipAmplitude}) and tip frequency (\ref{sec:TipFreq}), for a good range of Cauchy numbers. Such results may not only contribute to the physics of the structural dynamics of plant canopies that exhibit coherent motion such as \textit{honami} and \textit{monami} but also, to a novel kind of \textit{Fluid-Structure Interactions} of a slender flexible cantilevered-element and vortices in an otherwise steady flow.

\section{Materials, set-up and methods}
\label{sec:set-up}
\begin{figure}
\begin{center}
\epsfig{file=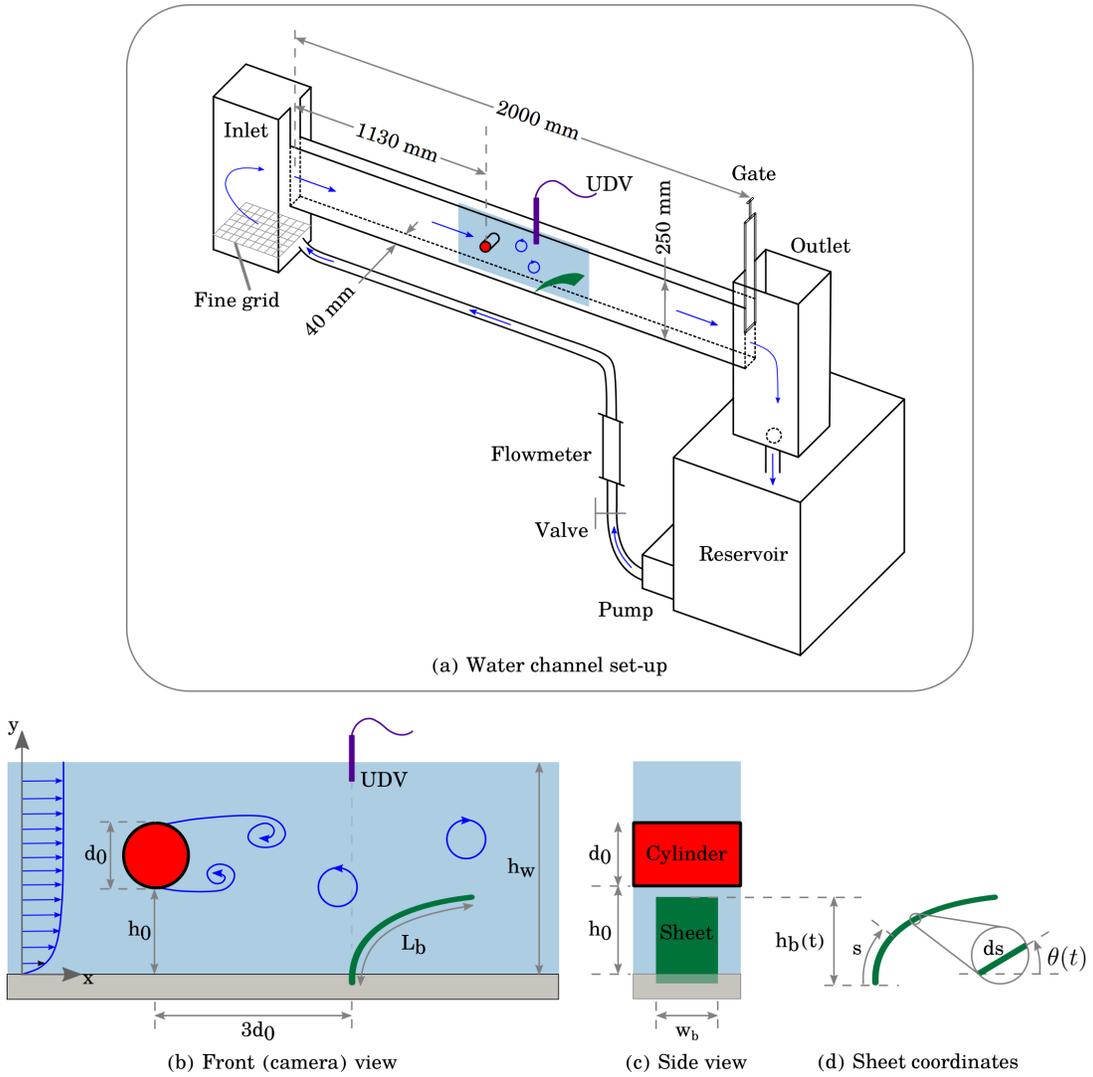,width=0.8\textwidth,keepaspectratio=true}
\end{center}
\caption{Schematic view of (a) the $2$-meter long water channel along with (b) top and (c) front view showing how the cylindrical obstacle is set-up in order to impose a vortex street forcing on a flexible sheet downstream.}
\label{fig:SchemaManip}
\end{figure}
All experiments are performed in a long-narrow water channel as schematized in Figure \ref{fig:SchemaManip} (a). Water from the pump ($700$ -- $2800$ lit/h) passes through a fine grid in the inlet and flows out in to a narrow $2$-meter long channel of width $4$ cm and height $25$ cm. By properly adjusting the outlet gate and the horizontal slope of the channel, it is possible to maintain a free-surface flow of uniform water height across the entire channel. At a little more than $1$-meter, a circular cylinder of diameter $d_0$ is fixed with its axis parallel to the floor, but perpendicular to the flow. Behind the cylinder (downstream), a thin flexible rectangular sheet is fixed firmly to the channel bottom. The sheet's foot is fixed at a distance $3d_0$ from the cylinder's center so that a well-developed vortex street is set-up \cite{Roshko1954_modRe, Szepessy1992_AR} in front of the sheet while the latter does not influence the cylinder's wake flow. Finally, to avoid the effect of free-surface on the Vortex-Forced-Oscillations (VfO) of the sheet, and also for the sake of simplicity, the cylinder's vertical location is taken as $h_0$ equal to the sheet height in a steady, uniform channel flow whose water height is kept identical throughout this work ($h_w = 22.1$~cm).  Figures \ref{fig:SchemaManip} (b) -- (c) illustrate this set-up wherein the cylinder and the sheet are exposed to an uniform channel flow directly orthogonal to the sheet's section $L_b w_b$.

In the range of depth-averaged flow speed $U_0 = 2.2$ -- $8.8$ cm~s$^{-1}$ and diameters $d_0 = 10$, $20$, $40$ mm used here, the cylinder Reynolds number $Re_d = U_0 d_0/ \nu$ varies over a decade, between $236$ and $3773$. An Ultrasonic Doppler Velocimetry (UDV) is used to obtain vortex street characteristics at the center plane of the channel and at a fixed position downstream, equal to $3$ times the cylinder diameter. The probe measures the instantaneous vertical velocity component in the water flow at an acquisition frequency of about $2$ MHz during $3$ minutes. The range of shedding frequency varies between $0.14$ and $2.13$ Hz as indicated in the Table \ref{tab:VortexCharacterics} (see also, Supplementary Material).
\begin{table}
  \begin{center}
  \begin{tabular}{c|c|c|c}
  		\hline
        {} 		&{} 		&{}					&{}		\\
      %{Diameter}	&{Cylinder Reynolds number}	&{Vortex shedding frequency}\\[3pt]
      	{$d_0$}		&{$U_0 d_0 / \nu$} 			&{$f_v$}				&{$f_v/f_n$}\\[3pt]
      	{(mm)}		&{} 		&{(Hz)}					&{}\\
         \hline
        {} 		&{} 		&{}					&{}		\\
      	{$10$ } 	&{$236$ -- $943$}			&{$0.14$ -- $0.63$}		&{$2$ -- $360$}\\
      	{$20$ } 	&{$472$ -- $1886$}			&{$0.28$ -- $1.26$ }	&{$1$ -- $210$}\\
      	{$40$ } 	&{$943$ -- $3773$}			&{$0.37$ -- $2.13$ }	&{$0.8$ -- $105$}\\
		{} 		&{} 		&{}					&{}	\\	
       	\hline

  \end{tabular}
  \caption{Characteristics of B\'enard-K\`arm\`an vortices for a range of water speed $U_0 = 2.2$ -- $8.8$ cm~s$^{-1}$. The shedding frequency were obtained using Ultrasonic Doppler Velocimetry (UDV). As compared to the cylinder Reynolds number $Re_{d} = U_0 d_0 / \nu$, the hydraulic Reynolds number $Re_{h} = U_0 D_h / \nu$ varies from $1730$ to $6920$, wherein the hydraulic diameter is $D_h = {4 h_w w_c}/{(2h_w + w_c)}$ and the water height $h_w = 22.1$ cm is kept constant for all experiments. The \textit{reduced} frequency $u = f_v/f_n \gtrsim 1$ for all cases considered here and also, it increases almost as $C_y^{9/5}$ (not shown here).}
  \label{tab:VortexCharacterics}
  \end{center}
\end{table}

Five different cases of low-density polyethylene (density $\rho_b = 920$ kg~m$^{-3}$ and mass ratio $0.93$) sheets from GoodFellow Cambridge Limited are used in our experiments. A snap-off blade knife was used to cut large sheets to prepare long blades. This is a delicate task especially for longer and thinner sheets, as the process often leads to some local plastic deformation at its edges. In addition, some defects were initially present in the sheets. See Table \ref{tab:PhysicalProperties} for the sheet length $L_b$, the thickness $e_b$ and other physical properties. The sheet thickness is chosen to be always very small compared to all other dimensions. \jon{For the sake of simplicity, the gap between the sheet and the channel walls is taken to be narrow, so that the force exerted by the channel flow on the sheet is predominantly due to profile drag.} The Cauchy number
\begin{equation}
C_{y} = \dfrac{C_d \dfrac{1}{2} \rho U_0^2}{E}\left( \dfrac{ L_b^3 w_b}{I} \right),
\label{eq:CauchyNumber}
\end{equation}
% = 12 \left(\dfrac{C_d \frac{1}{2} \rho U^2}{E}\right) \left( \dfrac{ L_b^3}{e_b^3} \right),
which expresses the ratio of the flow drag and the elastic restoring forces, is varied in wide a range between $\mathcal{O}(1)$ and $10^5$. Here, $U_0$ is the depth-averaged flow speed, $E$ is the Young's modulus, the second moment of inertia for a thin flexible sheet is taken as $I = w_b e_b^3/12$, and  $C_d$ is the profile drag coefficient of a thin flat plate perpendicular to the flow direction. Based on measurements from our previous work, namely,  \citet[see fig.~$6$a]{Barsu2016}, $C_d = 6$ is taken throughout the present work. 

An estimate for the natural frequency of a fixed-free cantilever beam in a fluid can be found in classical texts such as \cite{Blevins2015formulas, SummerFredsoe_2006}:
\begin{equation}
	f_{ni} = a_{i}\sqrt{\dfrac{E I/w_b L_b^4 }{\rho_b e_b + {\pi C_M \rho w_b}/{4}}},
	\label{Eq:FreqNat}
\end{equation}
% = a_{i}\sqrt{\dfrac{E \left(e_b/L_b\right)^3}{12 \rho_b e_b L_b + {3\pi C_M \rho w_bL_b }}} ,
where $a_i$ is a non-dimensional constant with ($a_{1} = 0.56$ and $a_{2} = 3.5$ for the first-order and second-order natural frequency, resp.) and $C_M$ is the added mass coefficient. Note that this \textit{natural} frequency is usually attributed to beams in fluids that undergo small-amplitude vibrations and small deflections. Hence, such a frequency might not be relevant to long flexible sheets considered in this study since they exhibit large deflections. Also, these sheets are \textit{pre-tensioned} due to the presence of a mean flow drag.
\begin{table}[H]
  \begin{center}
  \begin{tabular}{l|c|c|c|c|c|c|c}
  	\hline
        			&{} 		&{} 		&{}					&{}				&{}			&{}  \\
%      {Sheet ID}	&{Length}		&{Thickness}	&{Young's Modulus}		&{Cauchy number} 	&{Buoyancy number}	&{Natural frequency}
%      \\[3pt]
       {ID} 			&{$L_b$} 		&{$e_b$} 		&{$E$}					&{$C_y = \dfrac{C_d \frac{1}{2} \rho U^2}{E} \left( \frac{ L_b^3 w_b}{I} \right)$}				&{$B_a = \dfrac{\Delta \rho g e_b}{E} \left( \frac{ L_b^3 w_b}{I} \right)$}			&{$f_{n1}$}  &{Symbol}\\
{}        			&{(mm)} 		&{(mm)} 		&{($\times 10^6$ Pa)}					&{}				&{}			&{ (Hz) }  &{}	\\
         \hline
        			&{} 		&{} 		&{}					&{}				&{}			&{}  	&{}  \\
       $S442$   	&{$84$ } 		&{$0.19$ }	&{$210$}		&{$5$ -- $88$}				&{$0.55$}		&{$0.286$}	&{$\triangleleft$ }\\
       $S1263$    	&{$240$ } 	&{$0.19$ }	&{$210$}			&{$130$ -- $2000$}			&{$12.8$}		&{$0.035$}	&{$\Box$}\\
       $S1400$    	&{$84$ } 		&{$0.06$ }	&{$230$}		&{$908$ -- $4200$}			&{$8.3$}		&{$0.042$}	&{$\bigcirc$}\\
       $S2000$    	&{$200$ } 	&{$0.10$ }	&{$250$}			&{$2100$ -- $9800$}			&{$32.1$}		&{$0.018$}	&{$\medwhitestar$}\\
       $S4000$   	&{$240$ } 	&{$0.06$ }	&{$230$}			&{$12500$ -- $98200$}		&{$192.4$}		&{$0.005$}	&{$\Diamond$}\\
        			&{} 		&{} 		&{}					&{}				&{}			&{}  \\
       \hline% 0.043717       1.0196      0.40027       1.7893       9.3357

  \end{tabular}
\caption{Geometric and mechanical properties of low-density polyethylene sheets along with the range of related Cauchy numbers attained in this work. All sheets are less denser than water (density $\rho_b = 920$ kg~m$^{-3}$) and their width ($w_b  = 15$ mm) is kept constant throughout this work. The sheet ID also indicates their stiffness ratio given by $L_b/e_b$.}
  \label{tab:PhysicalProperties}
  \end{center}
\end{table}
\begin{figure}
\begin{center}
\epsfig{file=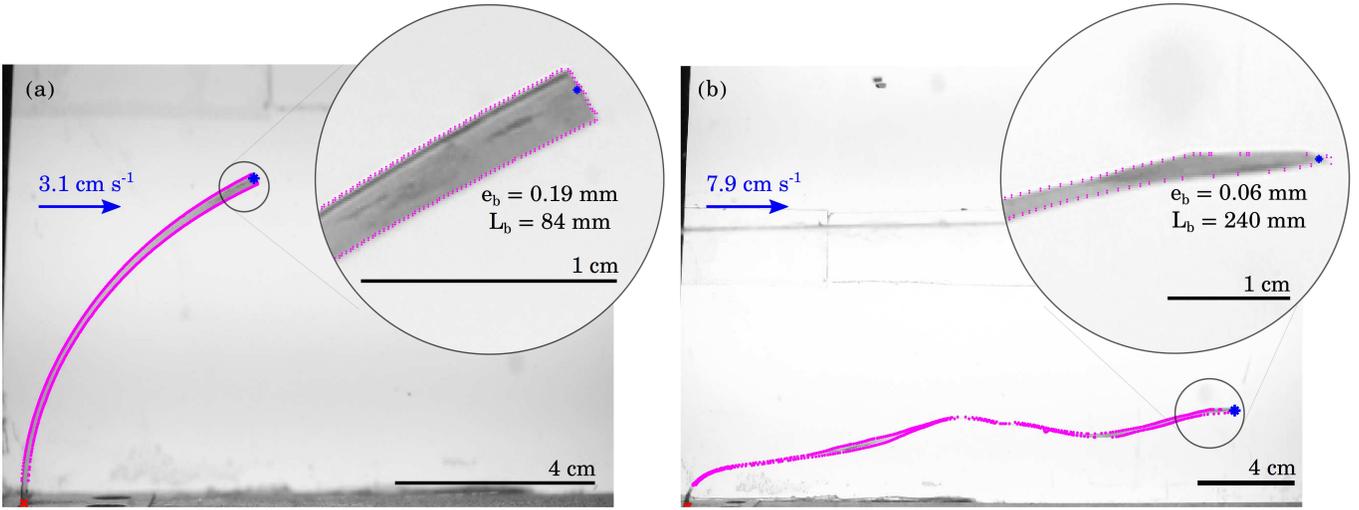,width=1\textwidth,keepaspectratio=true}
\end{center}
\caption{Two examples showing the results of sheet edge detection via \textit{ImageJ} with insets displaying a zoom on the tip of the sheet. (a) S$442$: $C_y = 11.3$, $Re_h = 2.5 \times 10^3$, $Re_d = 1.4 \times 10^3$ and (b) S$4000$: $C_y = 7.83 \times 10^4$, $Re_h = 6.2 \times 10^3$, $Re_d = 3.4 \times 10^3$. Here, dots (pink) show the detected sheet edge while an asterisk (blue) and a cross (red) each indicate the sheet's tip and foot, respectively. }
\label{fig:ImgDetec}
\end{figure}
A high-resolution, full-frame digital camera (Sony $\alpha 7$) is used to image the Vortex-forced-Oscillations (VfO) of the flexible sheet at a rate of $25$ images per second for a time period of $7$ to $8$ minutes (see supplementary videos). The resulting images are analysed using the open-source freeware \textit{ImageJ} \cite{Schindelin_NatureMethods2012fiji} and algorithms therein for brigthness thresholding \cite{Kapur_CompVision1985thresholding,Tsai_1985CompVisionthresholding}, edge detection, etc. Sample images and the corresponding edge detection (dots, pink) are shown in figures \ref{fig:ImgDetec}(a) and (b). Clearly, the contour of the sheet is well-detected. The sheet ``tip'' is taken as the center of the last identified sheet edge. This allows for a well-resolved tip detection amplitudes in the order of a few $10$-th of a millimetre.

In the following sections, all experimental results are presented in terms of the depth-averaged water speed across the time-averaged deflected sheet height $\bar{h}_b$. This speed is based on the classical \textit{Coles} law for the channel velocity profile $U(y)$, so that $U_h \bar{h}_b = \int_0^{\bar{h}_b} U(y) dy $.  Each symbols, namely, $\triangleleft$, $\Box$, $\bigcirc$, $\medwhitestar$, and $\Diamond$, represent different sheets of increasing stiffness ratio ($L_b/e_b$), respectively, provided in Table \ref{tab:PhysicalProperties}.

\section{Time-averaged sheet reconfiguration}
\label{sec:SheetReconfig}
%Related to the sheet tip dynamics is the mean sheet tip position. 
In the absence of a vortex street, at any chosen flow rate and constant water height, a sheet bends in the flow direction and exhibits a deflected height $h_b < L_b$. This leads to the well-known drag reduction since the profile drag force experienced by a flexible sheet $F_{d} = C_d \frac{1}{2} \rho U^2 h_b w_b$ is smaller than that experienced by the same sheet, if it was rigid (here, $C_d$ is the sheet drag coefficient). Such a drag reduction is often quantified by expressing the so-called \textit{static} reconfiguration number $\mathcal{R} = h_b/L_b$ as a function of the flow speed $U_0$ and the so-called Vogel number ~\cite{Vogel_1989} i.e, $h_b/L_b \sim U_0^{\mathcal{V}}$. Thereby, the profile drag can be written as a power law 
$$F_{d} \propto \dfrac{1}{2} \rho U_0^{2+\mathcal{V}} A_f,$$
where $A_f = L_b w_b$ is the undeformed sheet frontal area. Then, drag reduction by reconfiguration is observed if the Vogel number is $\mathcal{V} < 0$. For example, the Vogel number $\mathcal{V} = -2/3$ for long thin blades which are anchored to the flow bed at its one end~\cite{DeLangre2008}. While the reconfiguration number expresses the frontal area reduction, a more appropriate way to estimate drag reduction by reconfiguration is to compute the effective blade length proposed by \citep{Luhar_Reconfig2011}. Such a length also includes the streamlining effect due to bending at the foot of the sheet. For the sake of simplicity, we consider the reconfiguration number which accounts only area reduction. This provides an upper bound for drag reduction by reconfiguration as well \cite{Luhar_Reconfig2011}.

\begin{figure}
\begin{center}
\epsfig{file=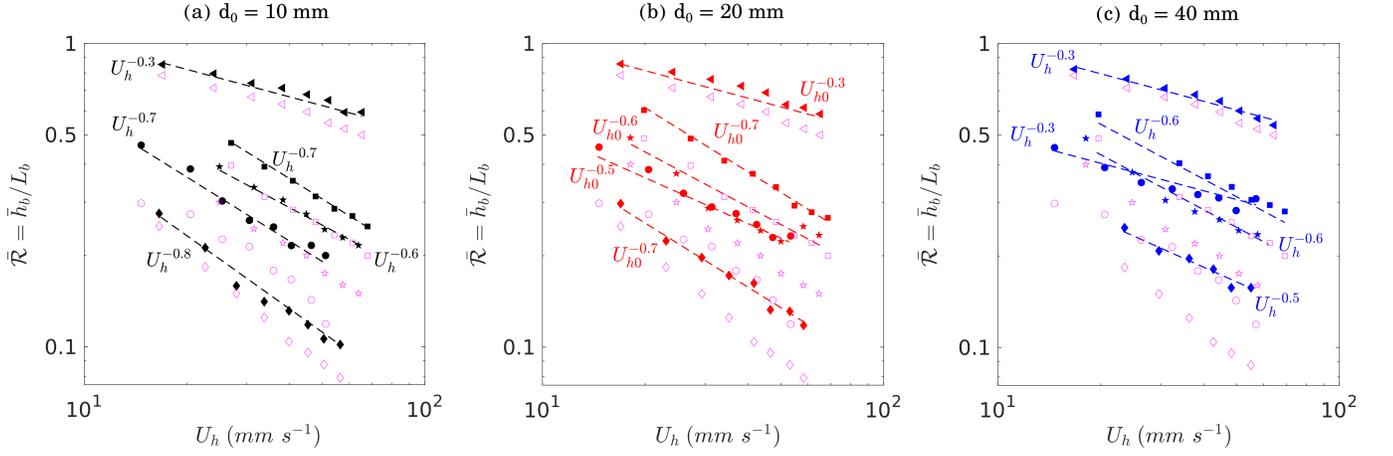,width=1\textwidth,keepaspectratio=true}
\end{center}
\caption{Time-averaged sheet reconfiguration $\bar{ \mathcal{R} }_b = \bar{h}_b/L_b $ in the presence of B\'enard-K\`arm\`an vortices as a function of the depth-averaged water speed across the sheet height ($U_h$). All symbols indicate different stiffness ratio ($L_b /e_b$) as given in Table II. Open symbols represent data for the same cases as filled symbols but in the absence of an upstream cylinder.}
\label{fig:Reconfig}
\end{figure}

In this context, Figure \ref{fig:Reconfig} displays the \textit{dynamic} reconfiguration number $\bar{\mathcal{R}} = {\bar{h}_b}/{L_b}$ against water speed for each of the five sheets in our study. Here, each column correspond to data for a particular cylinder diameter $d_0$. Reconfiguration data in the absence of any upstream cylinder is also presented in each figure for comparison (open symbols; light magenta). We observe that the average reconfiguration number is always lesser than unity and the related time-averaged sheet deflection can be as small as one-tenth of its length. Furthermore, in all cases corresponding to a fixed $L_b/e_b$, $\bar{\mathcal{R}}$ decreases when the flow speed increases. So, the frontal area of a given sheet that is exposed to the incoming flow decreases with speed, leading to an average drag-reduction by reconfiguration. By comparing each filled symbol in Figure \ref{fig:Reconfig} with its corresponding open symbol, we note that $\bar{\mathcal{R}} = {\bar{h}_b}/{L_b}$ always falls above the values of ${\mathcal{R}_0} = {{h}_0}/{L_b}$. So, the time-averaged sheet-deflection at a given speed is systematically bigger than its counter-part in the case of a flexible sheet free of an upstream cylinder. We refer to this as a ``lift-up'' effect. It is observed to be more pronounced for bigger cylinder $d_0$ and also, at higher water speeds $U_h$. However, at a fixed $d_0$ and $U_h$, the influence of sheet's stiffness ratio $L_b/e_b$ on the ``lift-up'' effect is not straight-forward.

\jon{Direct measurement of drag force is not possible in our experimental set-up. However, a measure of the average drag-reduction during the vortex-forced-motion of thin flexible sheets can be inferred from Figure \ref{fig:Reconfig} by estimating the so-called Vogel number $\mathcal{V}$.} As it is common for the case of a steady flow over a flexible sheet, we associate a power law $\bar{h}_b/L_b \sim U_h^{\mathcal{V}}$ and deduce a Vogel number $\mathcal{V}$ for each stiffness ratio ($L_b/e_b$). Except for the sheet with the smallest stiffness, all the other sheets present a Vogel number $\mathcal{V} \approx -0.6 \pm 0.1$. Note that similar values were previously obtained in the same water channel, but for submerged artificial canopies of kevlar sheets undergoing \textit{static} reconfiguration \cite[, fig $4$(a)]{Barsu2016}. We observe that the Vogel parameter is almost independent of cylinder diameter $d_0$ for sheets S$442$, S$1263$ and S$1400$. For more flexible blades (S$2000$ and S$4000$), a decrease in the Vogel parameter with increasing $d_0$ is observed. This suggests that vortices perhaps hinder the time-averaged drag reduction by reconfiguration of flow-bent flexible structures. In addition, this effect seems to be stronger for more flexible blades and bigger vortices.

\begin{figure}
\begin{center}
\epsfig{file=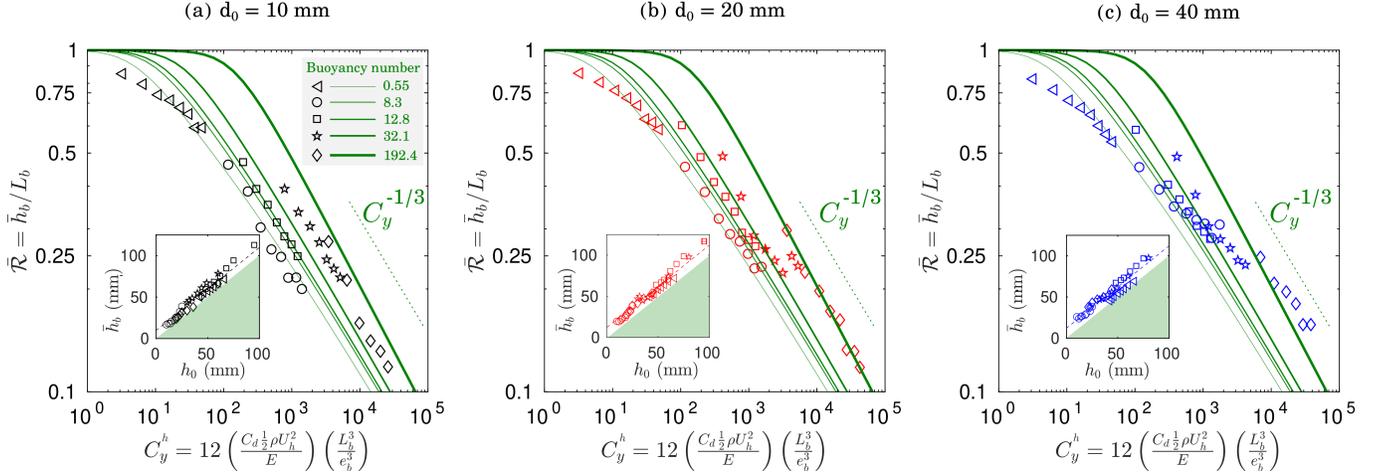,width=1\textwidth,keepaspectratio=true}
\end{center}
\caption{Time-averaged sheet reconfiguration $\bar{ \mathcal{R} }_b = \bar{h}_b/L_b $ expressed in terms of the \textit{local} Cauchy number $C_y^{h} = 12 \left({C_d \frac{1}{2} \rho U_h^2}/{E}\right) \left( { L_b^3}/{e_b^3}\right)$ and compared with continuous lines (green) as obtained from the bending beam model (eqn. \ref{eq:beamAdim}). The inset compares the average sheet deflection against the corresponding cylinder's vertical position.}
\label{fig:ReconfigCY}
\end{figure}
It is now well-established that a bending beam, accounting for the \textit{local} flow drag and the Archimedes force due to buoyancy provides a satisfactory mechanical model for \textit{static} reconfiguration of flexible sheets \cite{Alben2002, Luhar_Reconfig2011, Barsu2016, Leclercq_JFS2016}. If $s$ is the curvilinear coordinate along the sheet and $\theta(s)$ the \textit{local} sheet deflection as represented in fig. \ref{fig:SchemaManip} (d), the restoring bending moment $\mathrm{M}(s)$ at any arbitrary distance $s$ from the sheet's foot should be given by
\begin{equation}
	{{EI}\dfrac{d\theta}{d s}} = \int_{s}^{L_b} {\left( x(\xi) - x(s) \right)} dF_{\mathrm{A}} -  \int_{s}^{L_b} {\left( \xi - s \right)} dF_\mathrm{D},
	\label{eq:beam}
\end{equation}
where $dF_{\mathrm{A}} = \Delta \rho g \left( e_b w_b d \xi \right)$ is the \textit{local} Archimedes force and $dF_\mathrm{D} = C_d \frac{1}{2} \rho U^2 \sin ^2{\left(\theta(\xi)\right)} dA_{f}$ is the normal component of the \textit{local} profile drag, with $\Delta \rho = \rho - \rho_b$ the density difference between the fluid and the sheet, and $dA_f = w_b d \xi$ the \textit{local} frontal area of the \textit{reconfigured} sheet\footnote{Note that the effect of the sheet's curvature on the bending moment due to the drag force and the tensile stress along the length of the sheet are neglected for simplicity.}. The above equation can be further simplified by taking both $U(s) \equiv U_0$ and $EI$ to be invariant across the sheet so that
\begin{equation}
\dfrac{d^3 \theta}{d \tilde{s}^3}  = B_a \left( \sin \theta \left(1 - \tilde{s}\right) \dfrac{d \theta}{d \tilde{s}} + \cos \theta \right) - C_y \sin^2 \theta,
\label{eq:beamAdim}
\end{equation}
where $\tilde{s}=s/L_b \in \left[ 0, 1 \right]$ is the non-dimensional curvilinear coordinate, $B_a = \Delta \rho g w_b e_b L_b^3/EI$ is the so-called Buoyancy number and $\mathcal{C}_{y}= C_{d}\rho w_b L_b^3{U}^2_0/2EI$ is the Cauchy number. The typical values for these non-dimensional numbers are also provided in Table \ref{tab:PhysicalProperties}. This model equation can be readily solved by applying the boundary conditions at the sheet extremities $\theta = \pi/2$ and $d\theta/ds = d^2\theta/ds^2 = 0$, for $\tilde{s} = 0$ and $\tilde{s} = 1$, respectively. 
\begin{figure}
\begin{center}
\epsfig{file=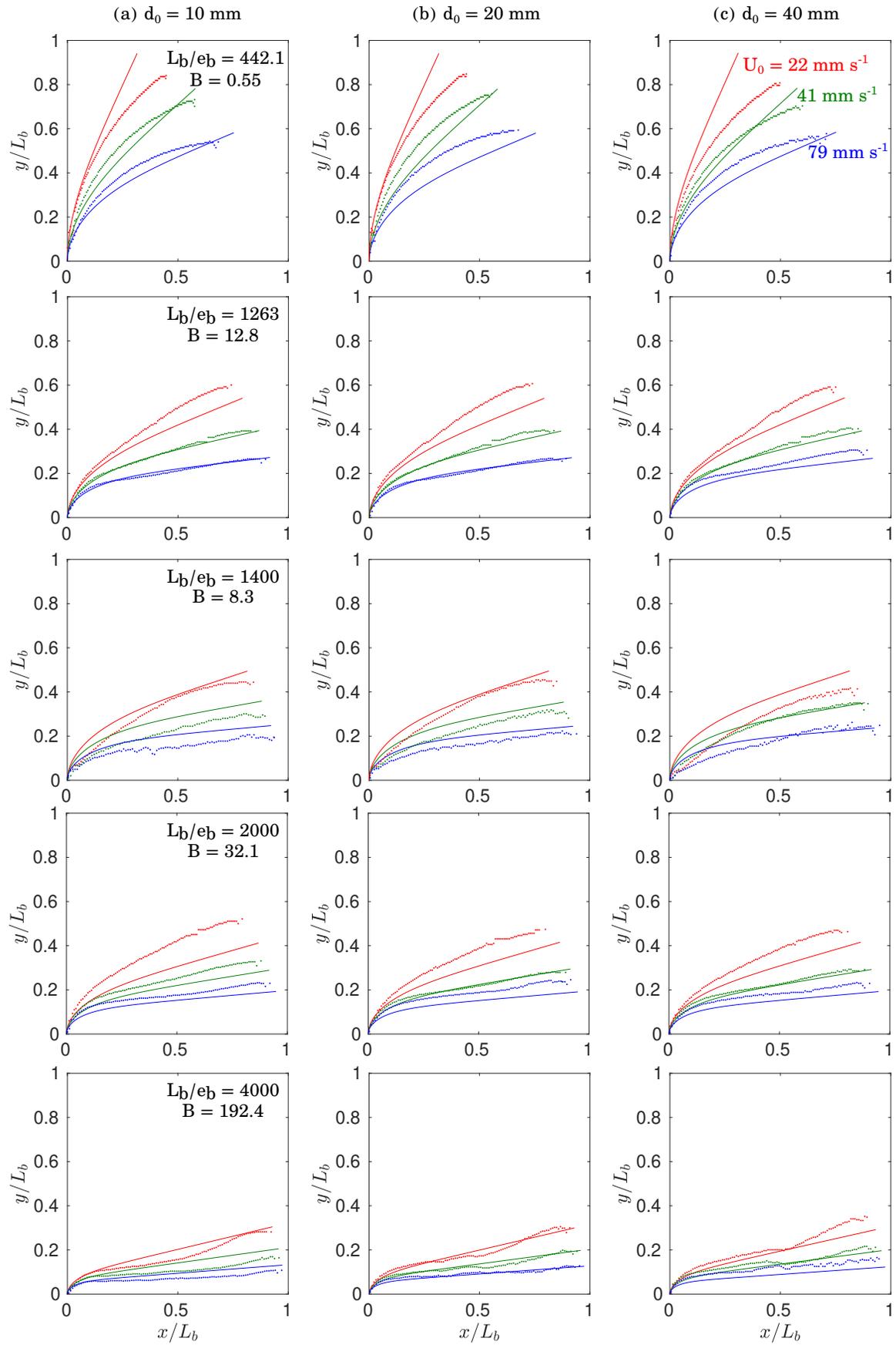,width=0.85\textwidth,keepaspectratio=true}
\end{center}
\caption{Typical time-averaged sheet shape from experiments (dots) as compared to bending beam model (continuous lines), allowing for the effect of buoyancy in a steady, uniform channel flow (eqn. \ref{eq:beamAdim}).}
\label{fig:ReconfigCompa}
\end{figure}

Let us now investigate the \textit{dynamic} reconfiguration number as a function of Cauchy number, as provided in Figure \ref{fig:ReconfigCY}. As before, we use the depth-averaged water speed $U_h$ across the time-averaged deflected sheet height $\bar{h}_b$ to express data in terms of the \textit{local} Cauchy number $C_y^{h} = 12 {C_d \frac{1}{2} \rho U_h^2}/{E} \left(L_b/e_b\right)^3$. Symbols represent experiments and continuous lines are computed from the bending beam model in eqn.~(\ref{eq:beamAdim}) at each various Cauchy and Buoyancy number. For a given blade, say for instance S$442$ as represented by $\triangleleft$, as $C_y$ increases, the sheet reconfiguration decreases monotonically. However, at a fixed Cauchy number $C_y$, reconfiguration data from all sheets do not collapse on a single master curve. A closer observation reveals that this is due to buoyancy which tends to increase the reconfiguration for sheets of larger Buoyancy number $B_a = 12 {\Delta \rho g e_b}/{E} \left({L_b}/{e_b}\right)^3$. Despite the fact that the above-mentioned bending beam model is only valid for the case of a \textit{static} reconfiguration under a steady uniform flow, it predicts the trend with the Cauchy number and also, the buoyancy number for all cases. It displays a reasonable agreement for $C_y > \mathcal{O}(1)$. 

Furthermore, figure \ref{fig:ReconfigCompa} compares the time-averaged sheet shape with that computed via the bending beam model (eqn. \ref{eq:beamAdim}). Note that the model provides a satisfactory estimate despite the strong approximations such as uniform, steady flow across the sheet. As already mentioned in Section \ref{sec:set-up}, defects were often present in the biaxially oriented thin polyethylene sheets. This is visible for the most flexible blade in our study, namely, S$4000$ at the smallest speed ($U_0 = 22$ mm s$^{-1}$, red color). Nevertheless, the influence of non-uniform velocity profile is perhaps more significant on discrepancies between the model and the experiments than any initial plastic deformation.

Finally, an estimation for these differences between experimental data and the steady flow model can be obtained by analysing the ``lift-up'' effect.  As observed in Figure \ref{fig:Reconfig} and Figure \ref{fig:ReconfigCY} (insets), the sheet deflection height $h_b/L_b$ in the absence of vortices is smaller than the \textit{dynamic} reconfiguration $\bar{ \mathcal{R} } = \bar{h}_b /L_b$. Also, the ``lift-up'' effect is more pronounced as the diameter of the upstream cylinder increases. And it arises from the presence of an upstream cylinder which leads to an oscillating velocity field in the vicinity of the sheet's tip, and a velocity defect in the longitudinal component of the mean flow downstream. So, for some average velocity defect $\bar{\delta u} < 0$, we deduce an estimate from the well-established power law ${h}_0/L_b \sim U_h^{\mathcal{V}}$ for \textit{static} reconfiguration to obtain  for its \textit{dynamic} counter-part as $\bar{ h }_b \sim h_0 \left( 1 + \mathcal{V} \bar{\delta u}_v/U_h \right)$. Suppose that this defect occurs in the sheet tip's neighbourhood of size $l_v$. Since the cylinder's vertical position is fixed at $h_0$ for each experiment, we then have $\bar{\delta u} \sim -U_h l_v/h_0$, upto a first-order correction of the mean flow $U_h$. Therefore, it is expected that $\bar{ h }_b \sim h_0 - \mathcal{V} l_v$, with $\mathcal{V} < 0$. Let us now compare this relationship with plots of $\bar{h}_b$ against $h_0$, as given in the insets of Figure \ref{fig:ReconfigCY}. We note that the general trend of data points reasonably agrees with the relation $\bar{ h }_b \sim h_0 - \mathcal{V} l_v$. Hence, the ``lift-up'' effect in \textit{dynamic} reconfiguration could be understood as a consequence of the presence of cylinder wake. It can also be inferred that the lengthscale $l_v$ depends on the cylinder diameter $d_0$. Nonetheless, the decrease in the Vogel exponent for more flexible sheets can not be completely explained via a small-velocity defect argument as here. This phenomenon is perhaps analogous to time-averaged reconfiguration in wave-induced oscillations of submerged flexible structures \cite{Luhar_JFS2016, Lei_CoastalEngg2019, Lei_JFS2019}. They obtain a scaling law for both the effective blade length \cite{Luhar_Reconfig2011} and the deflected height is found to be $\bar{h}_b/L_b \sim C_y^{-1/4}$ with a proportionality constant which depends on the ratio of the blade length and the wave amplitude. Thus, our results suggest that the \textit{dynamic} reconfiguration tends towards to wave-induced mean blade reconfiguration when the cylinder diameter increases.% While this brief discussion only qualitatively explains the major deviations from the above simplified model, a complete analysis of the flow-field near the tip is needed to quantitatively predict the \textit{dynamic} reconfiguration, which is beyond the scope of the present work.

\section{Sheet oscillation amplitude}
\label{sec:TipAmplitude}
\begin{figure}
\begin{center}
\epsfig{file=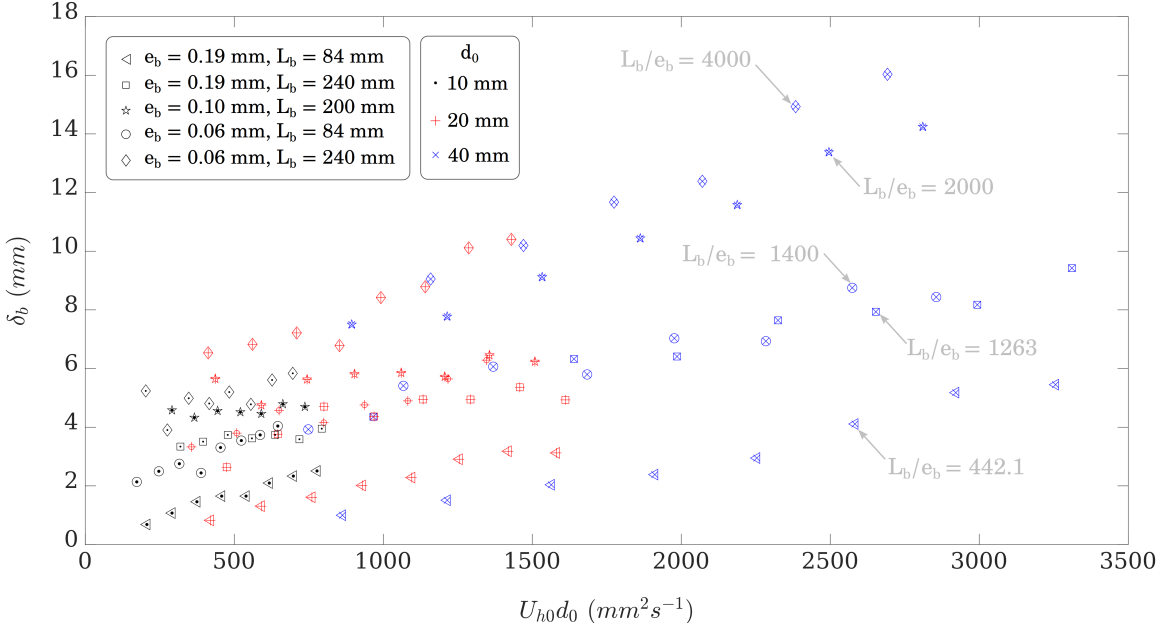,width=1\textwidth,keepaspectratio=true}
\end{center}
\caption{Tip oscillation amplitude $\delta_b$ as a function of an equivalent vortex \textit{circulation} $U_{h0} d_0$, where $U_{h0}$ is the water speed at the cylinder center.}
\label{fig:AmplitudeVsReynolds}
\end{figure}

If {$({x}_b(t), {y}_b(t))$} represent sheet tip's instantaneous position and {$(\bar{x}_b, \bar{y}_b)$} their time-average, then let us define the tip oscillation amplitude based on the standard deviation as in
\begin{eqnarray}		
	\delta_b \equiv 2\sqrt{\overline{ \left(x_b(t) - \bar{x}_b \right)^2} + \overline{\left( y_b(t) - \bar{y}_b \right)^2}},
	\label{eqn:TipAmp}	
\end{eqnarray}
where $\bar{\cdot}$ represents mean values. Figure \ref{fig:AmplitudeVsReynolds} presents such an amplitude as a function of the local water speed $U_{h0} = U(y = h_0)$  computed from the \textit{Coles law} channel velocity profile times the cylinder diameter $d_0$. Also, $U_{h0} d_{0}$ is proportional to the \textit{local} cylinder Reynolds number. Each open symbol corresponds to a stiffness ratio as given in Table \ref{tab:PhysicalProperties} and other symbols $\cdot$, {\color{red}$+$} and {\color{blue}$\times$} indicate cylinder diameters $d_0 = 10$, $20$ and $40$ mm, respectively. For a given sheet and cylinder, the tip fluctuation amplitude increases proportionally with the local water speed $U_{h0}$. Now, as the cylinder diameter is varied similar behaviors are again observed but the tip amplitudes $\delta_b$ are smaller for smaller diameters. In addition, sheets with higher stiffness ratio $L_b/e_b$ show increasingly larger amplitudes while data for sheets with approximatively same stiffness ratio, as in S$1263$ ($\Box$) and S$1400$ ($\bigcirc$) fall almost on the same linear trend line. 
In general, it is inferred from these observations that the tip amplitude not only increases with the sheet stiffness ratio $L_b/e_b$, but also with $U_{h0} d_{0}$.

\begin{figure}
\begin{center}
\epsfig{file=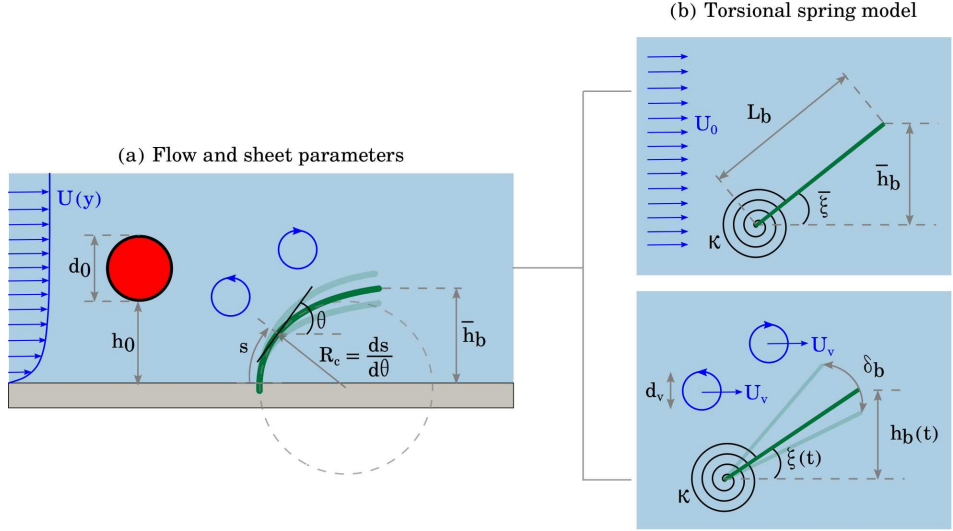,width=0.7\textwidth,keepaspectratio=true}
\end{center}
\caption{Schematic of the flow and sheet parameters along with that of the torsional spring model.}
\label{fig:TorsionModel}
\end{figure}

A simple model can be derived if we decompose the total work done by the vortex-laden flow on the flexible sheet into two distinct parts : $(i)$ the steady component of the flow leads to the average angular position, say $\bar{\xi}$, and $(ii)$ the periodic interaction between the vortices and spring-supported flat plate results in torsional vibrations of the spring, and hence the plate's angular position $\xi(t)$. Then, locally, the average profile drag-induced moment should be balanced by the average restoring moment in the deflected sheet $EI {d \theta}/{ds}$, where ${d \theta}/{ds}$ is the sheet's local curvature ($1/R_c$). As already observed in section \ref{sec:SheetReconfig}, at very large Cauchy number which compares drag force against the elastic restoring force, $\bar{h}_b \ll L_b$ and hence, it can be safely assumed that $EI {d \theta}/{ds} \approx {EI}/{\bar{h}_b}$, since $R_c$ is approximately equal to the time-averaged sheet deflection $\bar{h}_b$ (see Figure \ref{fig:TorsionModel}). And so, we obtain
\begin{eqnarray}
	\dfrac{EI}{\bar{h}_b}  &\sim &\left( C_d \dfrac{1}{2} \rho U_{0}^2 w_b \bar{h}_b \right) \times \bar{h}_b, \\ \notag
	\Rightarrow \dfrac{\bar{h}_b}{L_b} &\sim &C_y^{-1/3},
	\label{Eq:DynamicReconfig}
\end{eqnarray}
a result analogous to the well-known scaling for the \textit{static} reconfiguration number~\cite{DeLangre2008, Gosselin_JExpBot2019} that leads to drag reduction in flexible plates. When $C_y \ll 1$, on the other hand, ${\bar{h}_b} \approx {L_b}$. Note that the experimental data provided in the previous section, as in Figure \ref{fig:Reconfig}, match fairly well with the above large-Cauchy number scaling law, irrespective of Buoyancy number. In general, the above result could also be expressed as ${\bar{h}_b}/{L_b} \sim C_y^{\mathcal{V}/2}$, where the Vogel number $\mathcal{V} = -2/3$ at $C_y \gg 1$~\cite{DeLangre2008}.

Furthermore, we propose that the vibrational energy of the sheet is solely taken from B\'enard-K\`arm\`an vortices at some rate depending on the characteristics of the incoming unsteady flow, and during some time scale proportional to the shedding period $1/f_v$. Therefore, for the toy-model, we have
\begin{eqnarray}
	\dfrac{1}{2} \mathcal{K} \left(\dfrac{\delta_b}{L_b}\right)^2\sim \left(\dfrac{1}{2} \rho U_{0}^3 w_b d_v\right) \dfrac{1}{f_v},
	\label{Eq:BilanEnergie}
\end{eqnarray}
where the left-hand-side is the vibrational energy of the torsional spring and the right-hand-side is the product of \textit{local} kinetic energy transfer rate, taken here as proportional to $1/2 \rho U_{0}^3 w_b d_v$, and the typical timescale during which the transfer periodically takes place, i.e., $1/f_v$. Here, $d_{v}$ is some characteristic lengthscale of Bénard-Karman vortices. Also, in the above expression, the stiffness $\mathcal{K}$ of the \textit{pre-tensioned} torsional spring can be ascertained from the equilibrium condition that $\mathcal{K} \bar{\xi} \equiv EI {d \theta}/{ds} \approx {EI}/{\bar{h}_b}$, with $\bar{\xi} \approx {\bar{h}_b}/{L_b}$ when $Ca \gg 1$ since ${\bar{h}_b} \ll {L_b}$. Now, in terms of the Vogel number $\mathcal{V} = -2/3$ at $C_y \gg 1$, or $\mathcal{V} = 0$ at $C_y \ll 1$. By admitting that the \textit{dynamic} reconfiguration number is $\bar{h}_b/{L_b} \sim C_y^{\mathcal{V}/2}$, it can be deduced that $\mathcal{K} \sim \left({EI}/L_b\right) C_y^{-\mathcal{V}}$. Thereafter, the expression \ref{Eq:BilanEnergie} leads to
\begin{eqnarray}
	\left(\dfrac{\delta_b}{L_b}\right)^2 &\sim & \left( \dfrac{\rho U_{0}^2 w_b L_b}{EI} \right) \left( \dfrac{U_0 d_v}{f_v} \right) C_y^{\mathcal{V}}, \\ \Rightarrow \dfrac{\delta_b}{d_v} &\sim &C_y^{n} {St_v}^{-1/2},
	\label{Eq:AmpTipScaling}
\end{eqnarray}
where $St_v = f_v d_v /U_{0}$ is the Strouhal number based on a typical size of the eddies, and $n \equiv \left(1 + \mathcal{V}\right)/2 = 1/6$, or respectively $1/2$, for sufficiently large Cauchy numbers $C_y \gg 1$, or otherwise.

\begin{figure}
\begin{center}
\epsfig{file=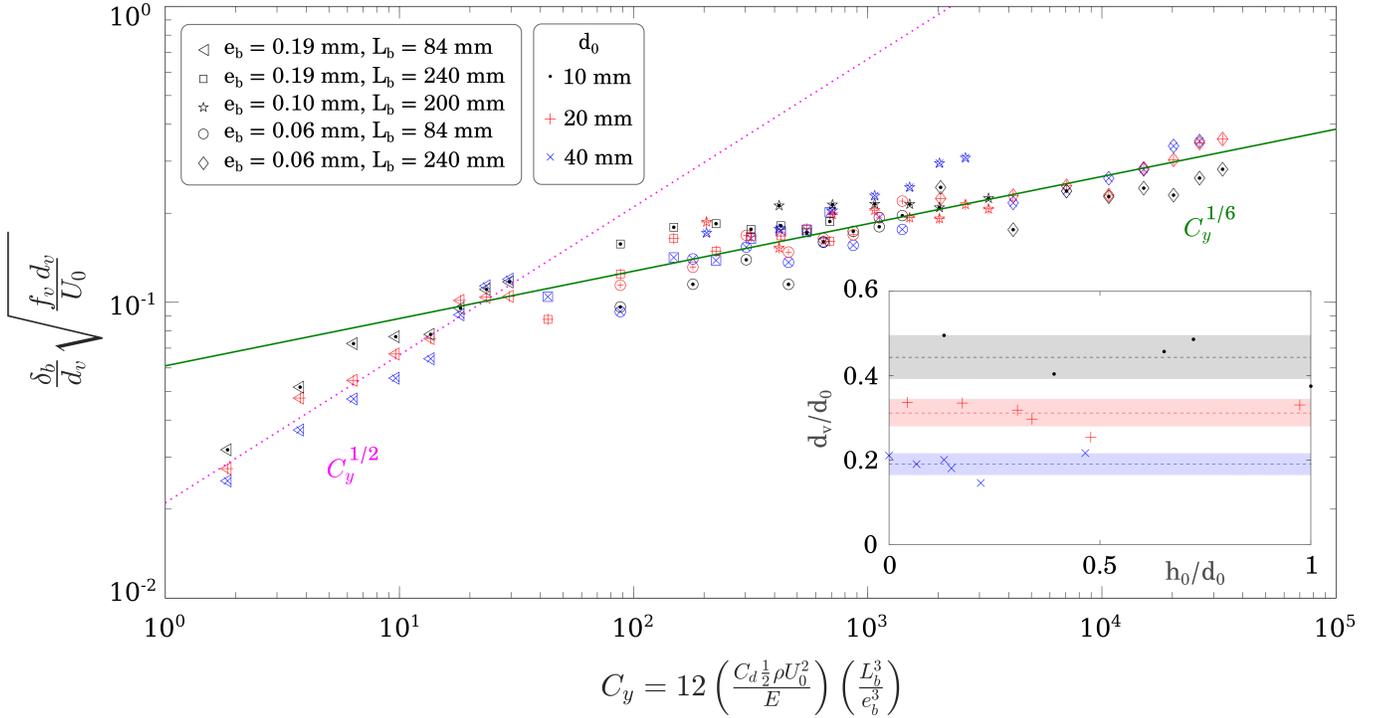,width=1\textwidth,keepaspectratio=true}
\end{center}
\caption{All data from Figure \ref{fig:AmplitudeVsReynolds} are expressed in terms of the \textit{rescaled} oscillation amplitude of the sheet's free-end $\delta_b /d_0$ as given by the relationship~\ref{Eq:AmpTipScaling} versus the Cauchy number $C_y = 12 \left({C_d \frac{1}{2} \rho U_0^2}/{E}\right) \left({ L_b^3}/{e_b^3} \right)$. Two distinct regimes are visible here.}
\label{fig:AmplitudeVsCyBis}
\end{figure}

Figure \ref{fig:AmplitudeVsCyBis} presents the same tip amplitude data given in Figure \ref{fig:AmplitudeVsReynolds} but instead, they are rescaled in terms of the experimentally obtained values of the Strouhal number $St_h$, cylinder centreline velocity $U_{h0}$ and the characteristic vortex length scale $d_v$ (see supplementary material and also, inset Figure \ref{fig:AmplitudeVsCyBis}). Clearly, experimental data ranging over all Cauchy numbers investigated here are regrouped around two distinct trend lines, namely $C_y^{1/2}$ and $C_y^{1/6}$, each corresponding to the case of moderately small and large Cauchy numbers, respectively, as expressed by the relation \ref{Eq:AmpTipScaling}. In the former, the sheet reconfiguration is sufficiently small and so, it represents vibration of a rigid sheet forced by a vortex street. Whereas, in the latter case, sheets can be considered to be flexible. These results strongly suggest that the torsional spring model contains the essential mechanism to explain the observed vortex-forced-vibration of thin sheets. They also imply that, analogous to drag-reduction via reconfiguration under an external flow, a flexible sheet experiences a smaller vibration amplitude compared to that of a rigid sheet when excited by a B\'enard-K\`arm\`an vortex street.  Finally, it is pointed out that the expression the \textit{dynamic} reconfiguration number may not in general be given by $\bar{h}_b/{L_b} \sim C_y^{\mathcal{V}/2}$ and this expression is valid only for either sufficiently small, or large, Cauchy numbers \cite{Gosselin_JFM2010}. In addition, the reconfiguration number should also depend on the Buoyancy number for intermediate Cauchy numbers \cite{Luhar_Reconfig2011}. So, it is possible that the transition between the two regimes of vortex-forced-oscillations may perhaps not be as abrupt as in our experiments.

\section{Sheet beating frequency}
\label{sec:TipFreq}
\begin{figure}
\begin{center}
\epsfig{file=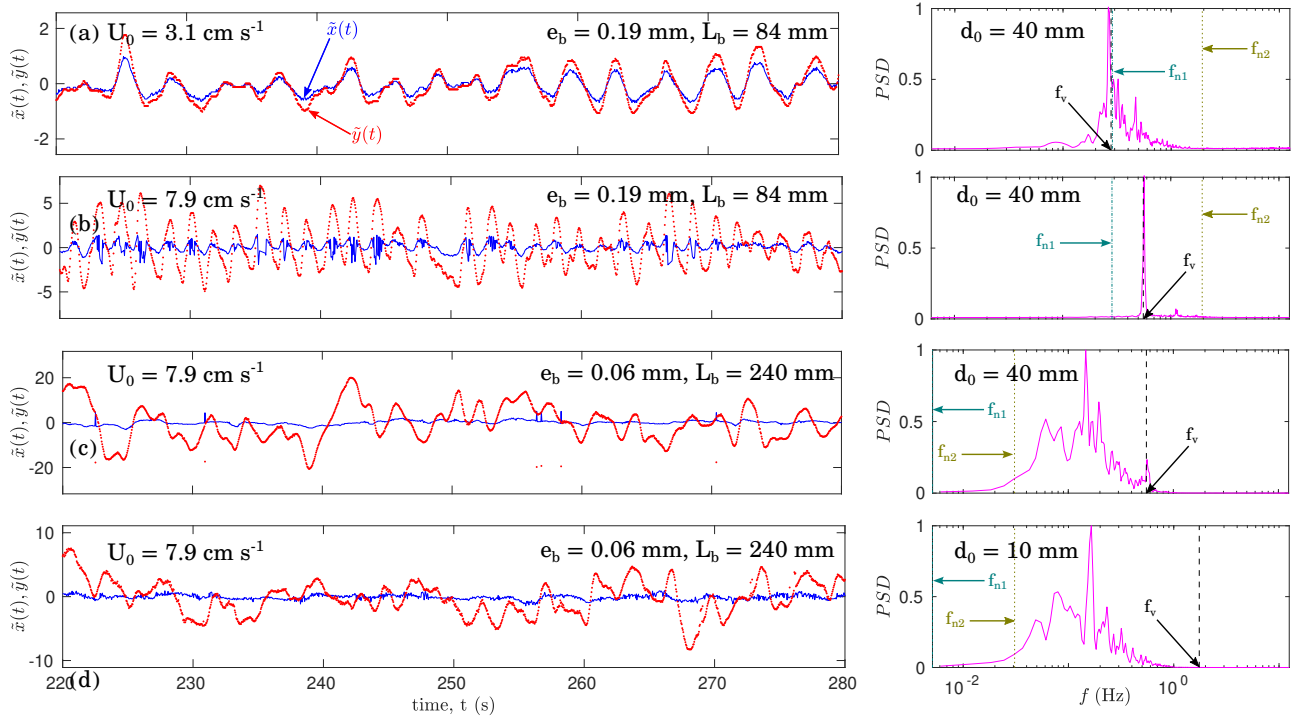,width=0.95\textwidth,keepaspectratio=true}
\end{center}
\caption{[\textit{Left}] Blade tip oscillations (in mm) of the ``stiffest'' sheet S$442$ at different depth-averaged water speed (a) $U_0 = 3.1$ cm~s$^{-1}$, (b) $U_0 = 7.9$ cm~s$^{-1}$ as compared with the ``most flexible'' sheet S$4000$ at (c) $U_0 = 7.9$ cm~s$^{-1}$ due to vortices shed by a cylinder of diameter $d_0 = 40$ mm. (d) Same as (c) but for $d_0 = 10$ mm. [\textit{Right}] Power spectral density of the sheet tip's vertical fluctuations. Arrows indicate various frequencies, namely, the vortex-shedding frequency ($f_v$) and the first two sheet natural frequencies ($f_{n1}$ and $f_{n2}$).}
\label{fig:BTip}
\end{figure}

The time evolution of tip response of a sheet S$442$ subject to B\'enard-K\`arm\`an vortices shed by a cylinder ($d_0 = 40$ mm) at the lowest and highest water speed is considered in this section. , respectively are displayed in Figures \ref{fig:BTip}(a) \& (b). In Figures \ref{fig:BTip}(a) , at $U_0 = 3.1$ cm~s$^{-1}$, both vertical and horizontal oscillations are of the same order of magnitude and also, they are synchronous. As the speed increases to $U_0 = 7.9$ cm~s$^{-1}$, $y$-fluctuations present an almost proportionally-increased \textit{back-and-forth} amplitude while $x$-fluctuations are much smaller in magnitude. Clearly, $x$-tip detection seems to be not so robust for this case.
%Nonetheless, vertical sheet tip position exhibits a reasonably continuous evolution wherein its frequency of oscillation $f_b$ seems to be greater than the case with smaller water speed $U_0 = 3.1$ cm~s$^{-1}$. 
Their corresponding power spectral density are provided on the immediate right of the same figures. Here, arrows are used to indicate the forcing frequency (vortex shedding) $f_v = 0.56$~Hz and the first natural frequency $f_{n1} = 0.29$~Hz, respectively. Clearly, the sheet tip oscillates at the vortex shedding frequency for these two cases. 

The above examples correspond to the sheet for which the relevant Cauchy numbers ($C_y = 5$ -- $88$) are the smallest. In comparison, Figure \ref{fig:BTip}(c) gives the tip fluctuations for S$4000$, at one of the largest Cauchy numbers ($C_y = 9 \times 10^{4}$). This sheet exhibits large amplitude oscillations in the vertical direction as big as the cylinder diameter $d_0 = 40$~mm while the horizontal oscillations remain small ($\lesssim 4$ mm) as before. The corresponding spectra of vertical tip position presents a peak at $0.15$~Hz, along with a smaller second peak at the vortex shedding frequency $f_v = 0.55$~Hz. Qualitatively similar temporal characteristics are observed for the sheet tip when the diameter $d_0$ is decreased. For example, Figure \ref{fig:BTip}(d) provides a typical data at $d_0 = 10$~mm to be compared with its equivalent case at the same water speed and sheet physical properties given in Figure \ref{fig:BTip}(c). Firstly, the vertical oscillations are diminished almost in proportions to the diameter-ratio. Secondly, the power spectral density presents a peak neither at the vortex shedding frequency $f_v = 0.55$~Hz, nor at the sheet natural frequency $f_v = 5 \times 10^{-3}$~Hz. Instead, the peak occurs at almost the same frequency ($0.16$~Hz $< f_v$) as for the case with the larger diameter $d_0 = 40$~mm. In fact, this peak is much closer to the flat-plate shedding frequency based on the average-deflected-height of the sheet, i.e, $0.145 U_h/\bar{h}_b = 0.2$ for these cases. In summary, the peak-normalised power spectra suggests that the tip motion in the $y$ direction oscillates either at the vortex shedding frequency $f_v$ or at a different frequency lesser than $f_v$ as and when $d_0$ decreases, or $L_b/e_b$ increases. 

\begin{figure}
\begin{center}
\epsfig{file=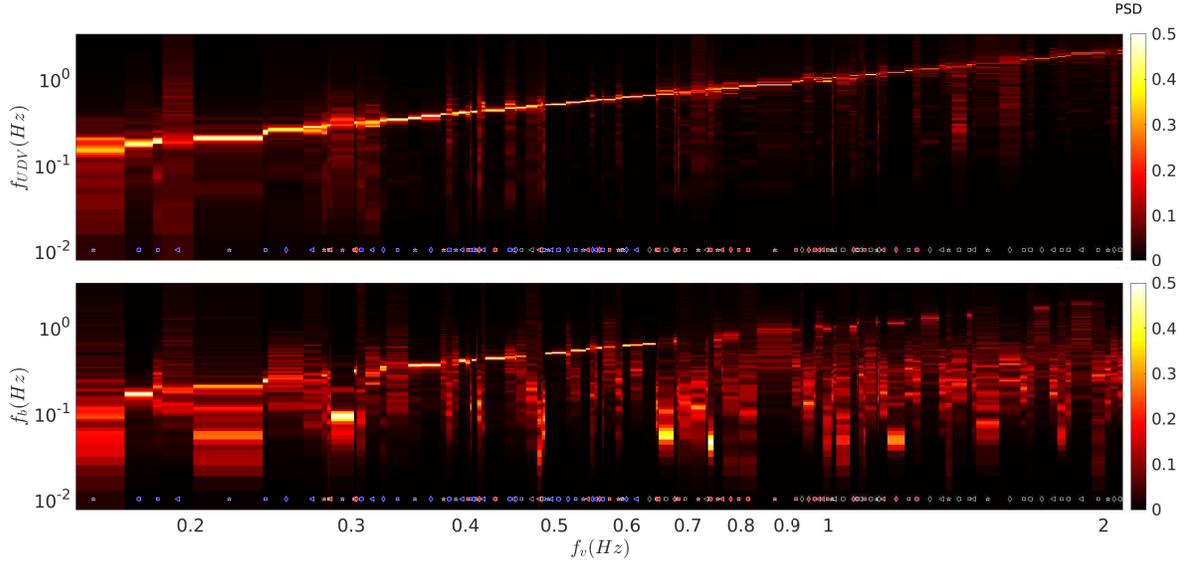,width=0.9\textwidth,keepaspectratio=true}
\end{center}
\caption{Power spectral density for all experimental cases as a function of the forcing frequency $f_v$, corresponding to imposed periodic shedding of B\'enard-K\`arm\`an vortices. The frequency content of the [TOP] vertical velocity behind the cylinder, as measured using an UDV and [BOTTOM] vertical fluctuations of the sheet's free-end. Dashed line is given simply to show the trend while symbols are provided for the sake of reference only (see Figure \ref{fig:AmplitudeVsCyBis}, for the corresponding sheet physical properties).}
\label{fig:PSD_FreqUDV}
\end{figure}

Figure \ref{fig:PSD_FreqUDV} (TOP) displays a color plot of the power spectral density of the UDV-measured, instantaneous vertical velocity at the cylinder mid-span and at a distance $3d_0$ downstream the cylinder wake.  As already mentioned, UDV measurements were acquired at $2$ MHz during $3$ minutes. Colors, from bright yellow to black, represent the normalised spectra of the $y$-component velocity field across the cylinder diameter. Here, the spectra at each $y$-coordinate ($y \in [h_0, h_0+d_0]$) is first computed and then an average in the discrete Fourier space is taken to obtain the normalised spectra. The latter presents a maximum (bright yellow, in Figure \ref{fig:PSD_FreqUDV} (TOP)) at a frequency $f$ which is, by definition, equal to the vortex shedding frequency $f_v$.
\begin{figure}
\begin{center}
\epsfig{file=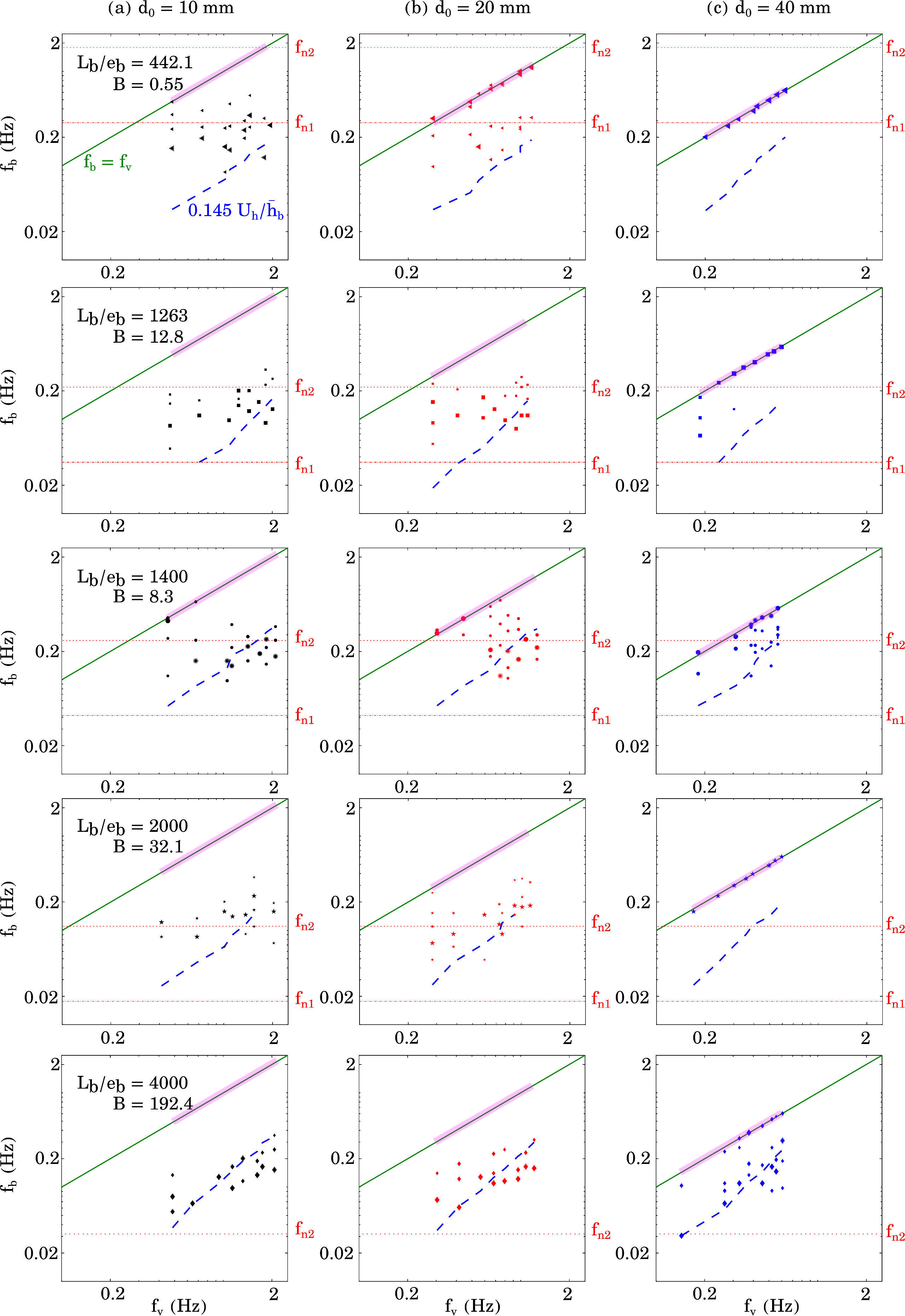,width=0.9\textwidth,keepaspectratio=true}
\end{center}
\caption{A detailed view of the measured blade peak frequency as a function of the vortex shedding frequency $f_v$. Symbols denote the same cases as in Figure \ref{fig:AmplitudeVsReynolds} while symbol size at each forcing frequency $f_v$ is proportional to the normalized power spectral density.  Here, the continuous line ({\color{OliveGreen}{------}}) represents $f_b = f_v$ while the dot-dash line ($\textcolor{red}{-\cdot-}$) and the dotted line ($\textcolor{red}{\cdots}$) indicate the sheet's first and second natural frequency $f_{n1}$ and $f_{n2}$, respectively, and ($\textcolor{blue}{- - -}$) is the flat-plate shedding frequency based on the sheet \textit{deflected} height $f_h = 0.145 U_h/\bar{h}_b$. Also, in each case, pink bands indicate the forcing frequency range.}
\label{fig:FreqUDVVsFreqBladeDetails}
\end{figure}

Similarly, Figure \ref{fig:PSD_FreqUDV} (BOTTOM) presents the power spectral density for the sheet-tip $y_b(t)$. For the sake of comparison, the $x$-axis is kept the same as just before. Therefore, this figure illustrates the energy content of the sheet-tip oscillation for each \textit{forcing} frequency, equal to the shedding frequency of the B\'enard-K\`arm\`an vortices $f_v$. Note that the sheet tip position spectra is not quite the same as the $y$-component velocity power spectra discussed just before. Clearly, there are many cases where the power spectrum is wide for a fixed $f_v$. It is thereby inferred that the sheet's vibrational energy is distributed across different frequencies and the dominant frequency $f_b$ varies from case to case. Nonetheless, the dominant frequency of sheet tip fluctuations is observed, in general, to be lower than the vortex shedding frequency $f_v$. Also, a second peak is often visible for some cases in Figure \ref{fig:PSD_FreqUDV} (BOTTOM).

To further elucidate the distribution of vibrational energy in the sheet-tip vertical motion, Figure \ref {fig:FreqUDVVsFreqBladeDetails} provides a few dominant tip frequencies $f_b$, i.e., those frequencies corresponding to the prominent peaks in the corresponding power spectral density, for various imposed vortex shedding frequency. Size of the data points are proportional to the normalized power spectra. In this figure, data in each row corresponds to a single sheet characteristics, in the ascending order of the sheet stiffness ratio $L_b/e_b$ while each column denotes data from experiments with the same cylinder diameter $d_0$. The cylinder diameter $d_0$ increases from left to right. Here, the continuous line (green) provided to indicate the cases when $f_b = f_v$, i.e., the cases where the observed beating frequency of the sheet is equal to the forcing frequency due to B\'enard-K\`arm\`an vortices.  Now it can be inferred from the data displayed in the last column, for experiments with the largest cylinder ($d_0 = 40$~mm), that the dominant frequency in the sheet-tip fluctuations is simply equal to the forcing frequency $f_v$, irrespective of the Cauchy number $C_y =  12 \left({C_d \frac{1}{2} \rho U^2}/{E}\right) \left({ L_b^3}/{e_b^3} \right)$. This is in contrast with the case when $d_0 = 10$~mm wherein $f_b < f_v$ for all sheets and multiple prominent peaks in the power spectra of the tip's vertical fluctuation $\tilde{y}_b (t)$. However, for some case, namely, S$442$ and S$1263$, the prominent frequencies are about the first, or second natural sheet frequency, respectively, as given by eqn.~(\ref{Eq:FreqNat}). Whereas, for the other cases, the data is scatter around the dashed-line (blue) which denotes the expression $f_h = 0.145 U_0/\bar{h}_b$, the flat-plate shedding frequency \cite{Blevins} based on the sheet deflected height $\bar{h}_b$, as known from the \textit{dynamic} sheet reconfiguration. We also observe similar dynamical regimes for the intermediate cylinder size $d_0 = 20$~mm : (i) For the most rigid sheet S$442$ and for $C_y < 20 $, the dominant blade frequency is seen to be the forcing frequency $f_v$, with a few dominant peaks either at the sheet's first natural frequency $f_{n1}$, (ii) For the most flexible sheets, namely, S$2000$ and S$4000$ and for $C_y > 10^3$ the sheets display oscillations at the vortex shedding frequency of an inclined flat plate given by $f_h = 0.145 U_0/\bar{h}_b$, and (iii) For moderate stiffness ratio and Cauchy numbers, the sheet oscillates either at one of its natural frequency which is close to the forcing frequency $f_v$ or at the inclined plate shedding frequency $f_h$. Finally, it is pointed out here that no conclusive evidence for a critical Cauchy number, nor a critical length scale for the vortices, is observed in these data. Nonetheless these observations strongly suggest that, in general, there is a transition from the forced-vortex-synchronous sheet-tip oscillation regime ($f_b = f_v$) to either a regime wherein the sheet-tip oscillations resemble the classical lock-in mode or a regime wherein the tip vibrations are induced by its wake characteristics. And the transition between these dynamical oscillation modes should depend on the relative size of the B\'enard-K\`arm\`an vortices and the stiffness ratio. 

\section{Conclusions}
\label{sec:conclusions}
%	Dynamic \textit{reconfiguration} and tip-dynamics of thin low-density polyethylene sheets subject to periodic forcing due to B\'enard-K\`arm\`an vortices in a $2$-meter long narrow water channel is investigated here.  
	The role of vortical structures in a transverse water flow on the flow-induced reconfiguration and vibration of a thin flexible sheet fixed to the floor at one end are experimentally investigated. Sheets of varying lengths and thicknesses ($L_b = 84$ -- $240$ mm; $e_b = 0.06$ -- $0.19$ mm) and three different cylinder diameters $d_0 = 10$, $20$ and $40$ mm are used for this purpose. Each experiment consist of rigidly anchoring the sheet to the channel bottom and then systematically exciting its free-end by vortices shed by a cylinder upstream. The forcing frequency is $f_v = 0.14$--$2.1$ $Hz$ for different depth-averaged water speed $U_0 = 22$ -- $88$~mm~s$^{-1}$, so that the cylinder Reynolds number varies between $Re_{d} \equiv \rho U_0 d_0/\mu = 240$ -- $3800$.
	
	Our experiments show that the time-averaged reconfiguration of a thin sheet follows qualitatively the same scaling with the Cauchy number $C_y = 12 \left({C_d \frac{1}{2} \rho U_{0}^2}/{E}\right) \left( { L_b^3 }/{e_b^3} \right)$ as in the case of a thin sheet in an uniform, steady flow bends, so that its time-averaged profile drag is reduced. A simple bending beam model for steady flow which takes into account the drag force, and also the buoyancy force, as in \cite{Luhar_Reconfig2011}, provides a reasonably good match with observations, if the flow speed $U_0$ is replaced by that of a local speed based on depth-averaged steady velocity profile given by \textit{Coles} law \cite{Coles1956law}. This suggests that the \textit{average} drag force $\bar{F}_d = C_d 1/2 \rho U_0^2 \bar{h}_b w_b \propto U_0^{2+\mathcal{V}}$. Here, $\bar{h}_b \propto L_b U_0^{\mathcal{V}}$ is the time-averaged sheet deflection and $\mathcal{V} < 0$ for drag reduction is the so-called Vogel number and it is observed to be $\mathcal{V} \approx -0.6 \pm 0.1$ in the present work. The influence of vortices is nonetheless visible for bigger cylinder diameters where there exists a slight ``lift-up'' effect leading to a small decrease in the sheet reconfiguration and Vogel exponent. This is attributed to the average velocity defect in the cylinder wake so that the mean sheet \textit{deflection} is given by $\bar{ h }_b \sim h_0 - \mathcal{V} l_v$, where $h_0$ is the deflection in the absence of vortices and $l_v$ is a characteristic lengthscale of the wake in the vicinity of the sheet tip.

	For a given blade thickness ($e_b$) and length ($l_b$), the oscillation amplitude ($\delta_b$) of the sheet tip increases with the Reynolds number $Re_{d}$. It is also demonstrated that the Cauchy number is the appropriate parameter to scale all data in terms of the non-dimensional amplitude $\delta_b/d_0$. The underlying mechanism that controls sheet-tip oscillations is then analyzed via a toy-model which consists of torsional spring-mounted rigid flat plate subject to an external flow which is decomposed into a uniform steady flow and a regular array of vortices. If the former is taken to control the \textit{average} sheet reconfiguration and the latter is assumed to provide the necessary work for the forced vibration the sheet. It is then shown that the rescaled oscillation amplitude $\delta_b/d_v \sim C_y^{(1+\mathcal{V})/2} St_v ^{-1/2}$, where $d_v$ is the typical lengthscale of the vortex core and $St_v = f_v d_v/U_{0}$ the related Strouhal number. In particular, for a relatively rigid sheet wherein the \textit{average} drag force provides only a very little sheet deflection  i.e., $\mathcal{V} \approx 0$, $\delta_b/d_0 \sim C_y^{1/2}$. This corresponds to the cases where the Cauchy number is either small, or at most moderate. Also, in this case, the sheet vibrates like a rigid curved plate as its local curvature does not change sign over the sheet's entire length. On the other hand, at large Cauchy number $C_y > 10^2$, for a relatively flexible sheet wherein the \textit{average} drag force results in a strong sheet reconfiguration i.e., when $\mathcal{V} \approx -2/3$ \cite{DeLangre2008}, the non-dimensional vibration amplitude of the sheet free-end $\delta_b/d_0 \sim C_y^{1/6}$. Note that, in the second regime, sheet exhibits modal oscillations like a flapping flag wherein transverse waves which travel forward towards the sheet free-end appear (see supplemental videos). 
	%The forward speed of such waves can attain as high as three times the \textit{local} flow speed at the cylinder centreline $U_{h0}$.
	
	In regards to the beating frequency ($f_b$) of the sheet free-end, three dynamical regimes are observed in this study: (i) Tip oscillations follow the forcing frequency $f_v$ corresponding to vortex shedding from the upstream cylinder, (ii) Tip oscillation frequency is related to vortex shedding behind a free inclined rigid plate of frontal height equal to the \textit{average} sheet deflection $\bar{h}_b$ such that $f_b \sim 0.145 U_{h}/\bar{h}_b$, and (iii) Sheet vibrations occur at one of its natural frequency $f_n$, near the forcing frequency. When the cylinder diameter $d_0 = 40$ mm,  sheet tip oscillates at the forcing frequency $f_v$, irrespective of the Cauchy number studied here ($C_y < 10^5$). For smaller cylinders, the sheet displays \textit{flow-induced vibration} controlled by its wake characteristics as in (ii) or a \textit{lock-in} motion at its natural frequency as in (iii). Furthermore, this transition is possibly dependent on the \textit{average} sheet reconfiguration and sheet thickness as well. 
	
	In a classical textbook by \citet{Blevins}, many investigations on Flow-Induced-Vibration are classified into two general categories depending on the incoming flow, namely, a steady or an unsteady flow. Our work is a sub-category of the latter case wherein we provide a case study for interactions between eddies and a flexible sheet. Indeed, more work is necessary to understand and hence predict the above-mentioned dynamical regimes via flow visualization, PIV measurements of the flow around the sheet and in its wake. \jon{The potential effect of sheet confinement is also left for future investigations.}%Moreover, the pertinence of these results for a canopy of flexible structures like plants (artifical or natural), and for different mass ratio as well, is left for future investigations.
	
\section*{Acknowledgements}
Authors thank St\'{e}phane Martinez from Universit\'{e} Claude-Bernard Lyon$1$ for his technical support building and maintaining the experimental set-up. PIV measurements are collected from an internship work by Christophe Lehmann.  We also acknowledge Emily M\"{a}usel and Cl\'{e}ment Pierrot-Minot (CP-M) for helping us measure some of the sheet's physical and mechanical properties. CP-M and JSJ thank Karine Bruy\`{e}re for her kind support and guidance with the linear-displacement facility (INSTRON \ $8802$) at Ifsttar-TS$2$, LBMC (Lyon-Bron) to estimate tensile strength of materials used in this work. This work had benefited from a joint French-German funding support, namely, the DFG-ANR project \textit{ESCaFlex} (ANR-$16$-CE$92$-$0020$, DFG grant $634058$).

%\bibliographystyle{unsrt}
%\bibliography{FlexiblePlatesUnderBvKvortices}

\end{document}

\title{Reconfiguration and oscillations of a vertical, cantilevered-sheet subject to vortex-shedding behind a cylinder.\\{\textsc{SUPPLEMENTAL MATERIAL}}}
\author{{J. John Soundar Jerome}$^1$, Yohann Bachelier$^1$, Delphine Doppler$^1$, \\Christophe Lehmann$^1$ and Nicolas Rivi\`ere$^2$}
\affiliation{$^1$ Univ Lyon, Universit\'{e} Claude Bernard Lyon $1$, Laboratoire de M\'{e}canique des Fluides et d'Acoustique, CNRS UMR--$5509$, Boulevard $11$ novembre $1918$, F--$69622$ Villeurbanne cedex, Lyon, France.}
\affiliation{$^2$ INSA de Lyon, Laboratoire de M\'{e}canique des Fluides et d'Acoustique, CNRS UMR--$5509$, Boulevard $11$ novembre $1918$, F--$69622$ Villeurbanne cedex, Lyon, France.}
%\date{}

\maketitle

In this supplemental material, we present movies of the experiments (\S 1), the vortex shedding frequency (\S 2) and finally, our method to estimate the typical vortex size for B\'enard-K\`arm\`an vortices (\S 3). Notations are the same as in the above-mentioned paper.

\section*{(1) Supplementary Movies} 
In the following, we describe the movies showing sheet reconfiguration and sheet tip oscillations.
\begin{itemize}
	\item Supplementary Video I: The effect of cylinder diameter, for the case of the most rigid blade (Length, $L_b = 84$mm and thickness $e_b = 180 \mu$m).
	\item Supplementary Video II: The effect of cylinder diameter, for the case of the most flexible blade (Length, $L_b = 240$mm and thickness $e_b = 63 \mu$m).
	\item Supplementary Video III: The effect of water speed, for two typical cases (S$442$ -- Length, $L_b = 84$mm and thickness $e_b = 180 \mu$m and S$2000$ -- Length, $L_b = 200$mm and thickness $e_b = 100 \mu$m).
\end{itemize}

\section*{(2) Vortex shedding frequency.} 
\begin{figure}[h]
\begin{center}
\epsfig{file=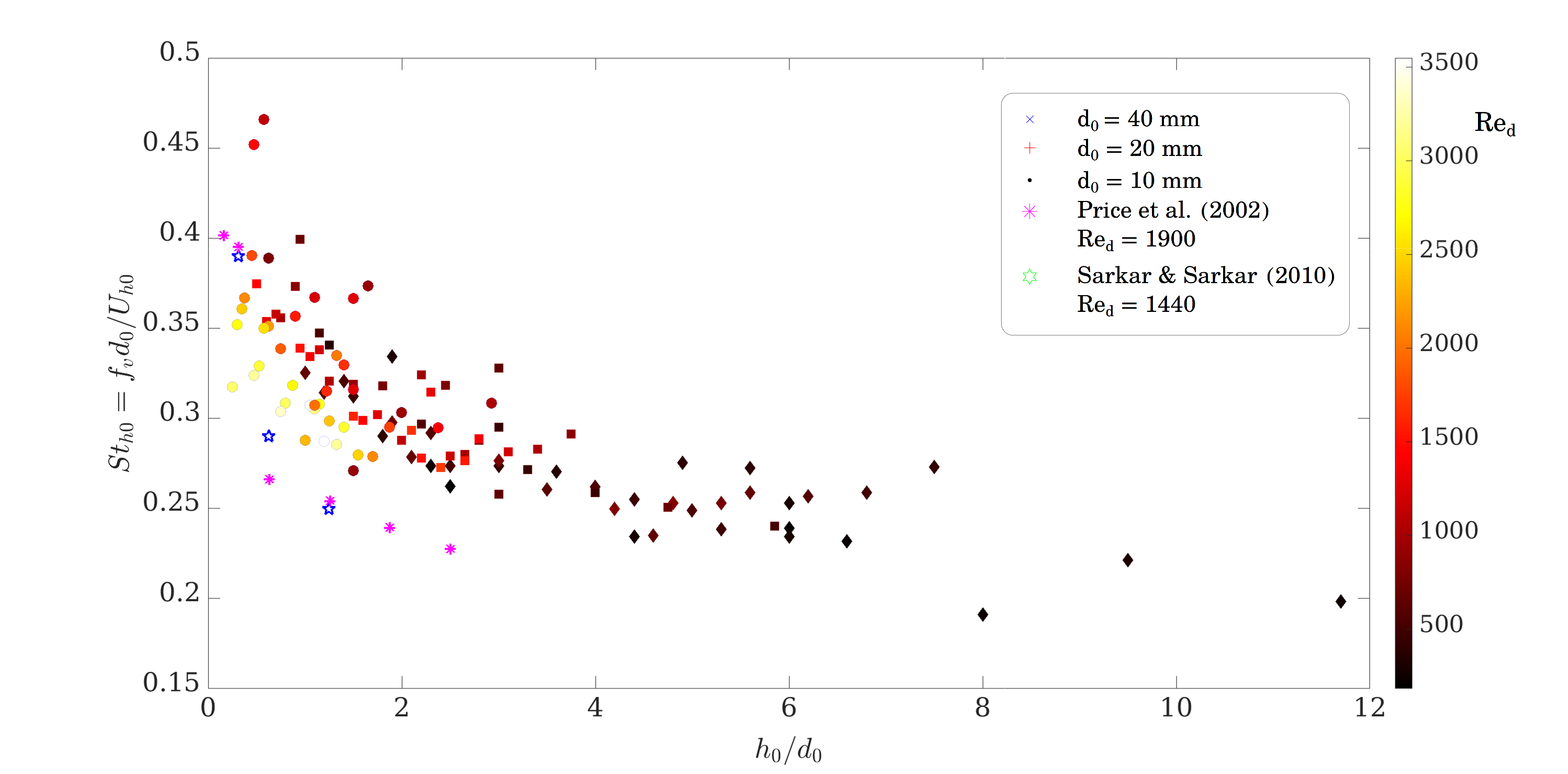,width=0.9\textwidth,keepaspectratio=true}
\end{center}
\caption{The \textit{modified} Strouhal number $St_{h0} = f_v d_0/U_{h0}$ based on cylinder center velocity $U_{h0}$ for the B\'enard-K\`arm\`an vortex shedding frequency $f_v$. }
\label{fig:FreqBrut}
\end{figure}

An Ultrasonic Doppler Velocimetry (UDV) is used to obtain vortex street characteristics at the center plane of the channel and at a fixed position downstream, equal to $3$ times the cylinder diameter. The probe measures the instantaneous vertical velocity component in the water flow at an acquisition frequency of about $2$ MHz during $3$ minutes. The range of shedding frequency varies between $0.14$ and $2.13$ Hz as indicated in Figure \ref{fig:FreqBrut}. A local velocity scale $U_{h0}$ is used for the Strouhal number. For this purpose, the mean channel flow profile $U(y)$ was computed computed using the so-called \textit{Coles law} \cite{Coles1956law, Kirkgoz_JHE1997velocity}. The \textit{modified} Strouhal number displays a general decreasing trend  similar to previous experimental \cite[{$*$}]{Angrilli_1982JFE, Price_JSFS2002} and $3$-D LES numerical investigations \cite[{$\star$}]{Sarkar_JFS2010}.
\begin{figure}[h]
\begin{center}
\epsfig{file=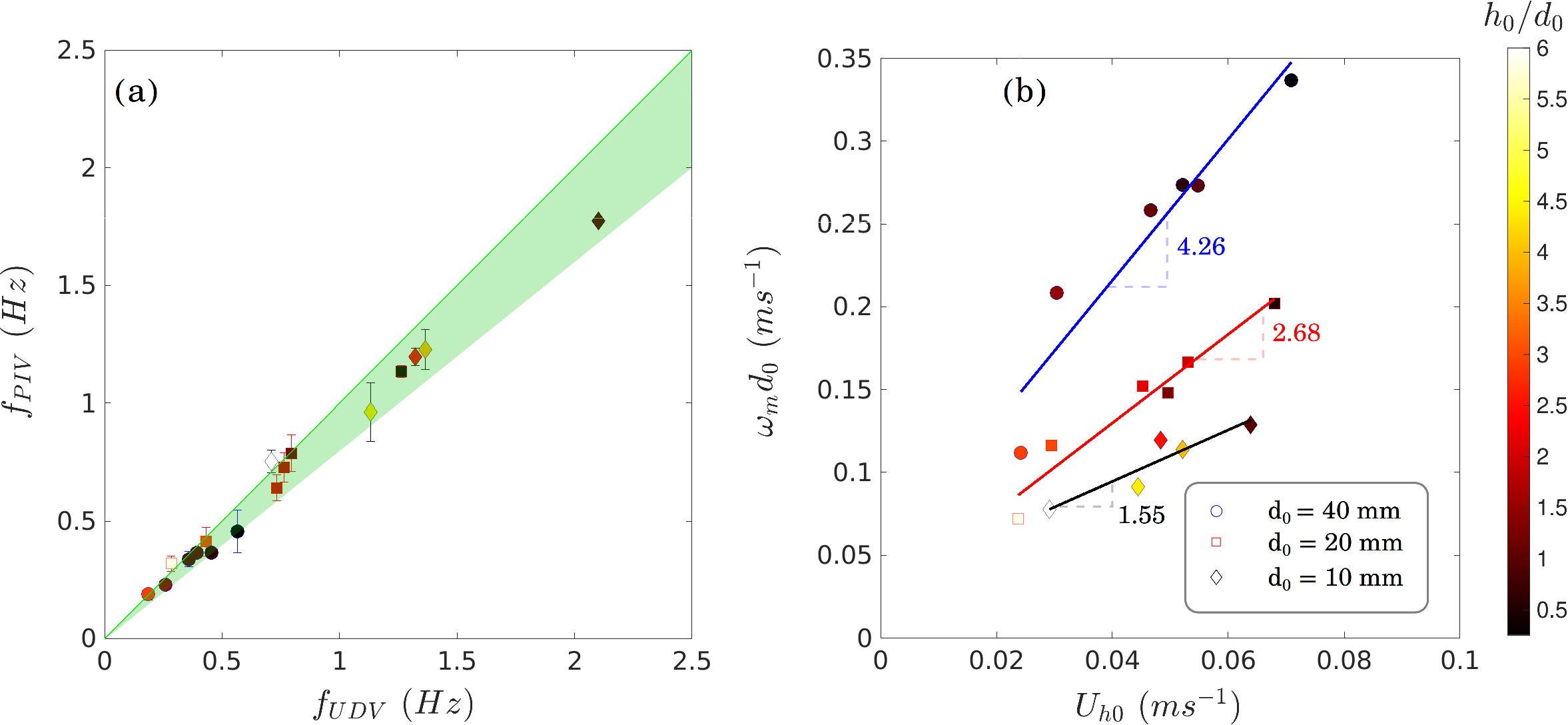,width=0.9\textwidth,keepaspectratio=true}
\end{center}
\caption{(a) Comparison between shedding frequency measured using Particle Image Velocimetry ($f_{PIV}$) and UDV ($f_{UDV}$) and (b) The product $\omega_m d_0$ against the cylinder centreline velocity $U_{h0}$ where \textit{average} maximum vorticity $\omega_m$ is obtained from the histogram of absolute maximum in the instantaneous vorticity profile at a fixed stream-wise location $x = 3 d_0$.}
\label{fig:PIV_vs_UDV}
\end{figure}

\section*{(3) Eddy size estimation.} 
For the purpose of this work, a set of PIV measurements in the region immediately downstream of the cylinder are undertaken. Not all configurations were considered but, in this study, only six pairs of ($h_0$, $U_0$) were chosen for each cylinder diameter $d_0$. Nonetheless, these parameters cover the wide range of values for $h_0$, $U_0$ and $d_0$ taken in our study. A high-speed camera with a resolution of $1024$ px $\times$ $1024$ px is used to capture images of a particle-seeded flow at a frame rate of $125$ fps. For the measurement, tracer particles with density $1005$ kg~m$^{-3}$ and a diameter of $50$ $\mu$m and $80$ $\mu$m are added to the flow. A system of mirrors scatters a LASER beam into a thin LASER sheet which, depending on the cylinder diameter $d_0$, covered from $6$ to $10$ times $d_0$. Standard recommendations \cite{Adrian_2011PIV} were followed for seeding, lighting and the relevant post-processing using \textit{DaVis Lavision} software. Finally, to obtain the frequency associated with the B\'enard-K\`arm\`an vortex street, a Fast Fourier Transformation is performed on the instantaneous vertical velocity given by PIV measurements. The frequency associated with the maximum spectral density is considered to be the vortex shedding frequency. To decrease the error, this process is repeated for each streamwise location on the centreline behind the cylinder to compute an average shedding frequency $f_{PIV}$. 
Such PIV measurements are compared with the UDV-measured shedding frequency $f_{UDV} = f_v$ in figure \ref{fig:PIV_vs_UDV}(a). The scatter plot also provides colored data points from black to bright-yellow which represent the ratio between $h_0/d_0 = 6$ and $h_0/d_0 = 0.25$, respectively. Clearly all data fall between the trend line $f_{PIV} = f_{UDV}$ and $f_{PIV} = 0.8f_{UDV}$. For larger frequencies, irrespective of $h_0/d_0$, the equality is less pronounced.

Furthermore, the instantaneous vorticity field can be computed from the measured velocity field at $x = 3d_0$ from the cylinder center. The absolute value of instantaneous vorticity  profile presents a maximum at a vertical position corresponding to either a counter-clockwise, or clockwise, rotating vortex. Now, at each time step, the absolute maximum vorticity including the sign is counted in order to build-up an histogram (not provided here). In general, such an histogram displays two peaks, each representing the most-likely maximum vorticity of the clockwise and counter-clockwise vortices. The half-distance between these peaks is then referred to be the \textit{average} maximum vorticity $\omega_m$ contained in shed vortices for a given set of experimental flow conditions, namely, $h_0$, $d_0$ and $U_0$. Figure \ref{fig:PIV_vs_UDV}(b) displays the product $\omega_m d_0$ as a function of the water speed at the cylinder centreline $U_{h0}$. Again, despite the variations in $h_0/d_0$, it is observed that for a given cylinder diameter, $\omega_m d_0 \propto U_{h0}$. So, by assuming that the maximum vorticity in B\'enard-K\`arm\`an vortices is $U_{h0}/d_v$ where $d_v$ is some typical size of the vortex core, it is then possible to note from figure \ref{fig:PIV_vs_UDV}(b), the ratio $d_v/d_0$ for different cylinder sizes $d_0$. 

In the main text, the typical eddy size $d_v$ is displayed as an insent to Figure 8. Here $d_v$ does not represent an average value as measured from the above data but it is taken as $U_{h0}/\omega_m$. 
\bibliographystyle{unsrt}
\bibliography{FlexiblePlatesUnderBvKvortices}